\documentclass[journal,12pt,onecolumn,draftclsnofoot,doublespace]{IEEEtran}

\usepackage{amsfonts}
\usepackage{amsmath}
\usepackage{graphicx}
\usepackage{color}
\usepackage{algorithm}
\usepackage{color}
\usepackage{algorithm,algorithmic}
\usepackage{enumitem}
\usepackage{array}
\usepackage{amssymb}
\usepackage{mathtools}
\usepackage{upgreek}
\usepackage{caption}
\usepackage{subcaption}
\usepackage{url}
\usepackage{cite}
\usepackage{amsmath}
\usepackage{eufrak}
\usepackage[autostyle]{csquotes}
\usepackage{verbatim}
\usepackage{amsmath,amssymb,amsthm}
\usepackage{tabularx}
\usepackage{enumerate}

\usepackage[utf8]{inputenc}
\usepackage[english]{babel}

\ifCLASSINFOpdf
\else
\fi
\begin{document}
%
\title{Coexistence Mechanism between eMBB and uRLLC in 5G Wireless Networks}



\author{Anupam Kumar Bairagi,~\IEEEmembership{Member,~IEEE,}
	Md. Shirajum Munir,~\IEEEmembership{Student Member,~IEEE,}
	Madyan Alsenwi,
	Nguyen H.~Tran,~\IEEEmembership{Senior Member,~IEEE,}\\
	Sultan S Alshamrani,
	Mehedi Masud,~\IEEEmembership{Senior Member,~IEEE,}\\
	Zhu Han,~\IEEEmembership{Fellow,~IEEE,}
	and~Choong Seon~Hong,~\IEEEmembership{Senior~Member,~IEEE}
	\thanks{Anupam Kumar Bairagi is with the Department of Computer Science and Engineering, Kyung Hee University, South Korea and Discipline of Computer Science and Engineering, Khulna University, Bangladesh (Email:anupam@khu.ac.kr).
		 
	Md. Shirajum Munir, Madyan Alsenwi, and ~Choong Seon~Hong are with the Department of Computer Science and Engineering, Kyung Hee University, South Korea (E-mail: \{munir,malsenwi,cshong\}@khu.ac.kr).

	Nguyen H.~Tran is with the School of Computer Science, The University of Sydney, Sydney, NSW 2006, Australia(E-mail: nguyen.tran@sydney.edu.au).
	
	Sultan S~Alshamrani is with the Department of Information Technology at the Taif University, Taif, KSA (E-mail: susamash@tu.edu.sa).

	Mehedi Masud is with the Department of Computer Science at the Taif University, Taif, KSA (E-mail: mmasud@tu.edu.sa).

	Zhu Han is with the Department of Computer Science and Engineering, Kyung Hee University, South Korea and Electrical and Computer Engineering Department, University of Houston, Houston, TX 77004, USA (Email: zhan2@uh.edu)

}
}


%

\markboth{IEEE Transactions on Communications}%
{Submitted paper}

\maketitle

\begin{abstract}
	\emph{uRLLC} and \emph{eMBB} are two influential services of the emerging 5G cellular network. Latency and reliability are major concerns for \emph{uRLLC} applications, whereas \emph{eMBB} services claim for the maximum data rates.  Owing to the trade-off among latency, reliability and spectral efficiency, sharing of radio resources between \emph{eMBB} and \emph{uRLLC} services, heads to a challenging scheduling dilemma. In this paper, we study the co-scheduling problem of \emph{eMBB} and \emph{uRLLC} traffic based upon the puncturing technique. Precisely, we formulate an optimization problem aiming to maximize the \emph{MEAR} of \emph{eMBB} UEs while fulfilling the provisions of the \emph{uRLLC} traffic. We decompose the original problem into two sub-problems, namely scheduling problem of \emph{eMBB} UEs and \emph{uRLLC} UEs while prevailing objective unchanged. Radio resources are scheduled among the eMBB UEs on a time slot basis, whereas it is handled for \emph{uRLLC} UEs on a mini-slot basis. Moreover, for resolving the scheduling issue of \emph{eMBB} UEs, we use \emph{PSUM} based algorithm, whereas the optimal \emph{TM} is adopted for solving the same problem of \emph{uRLLC} UEs. Furthermore, a heuristic algorithm is also provided to solve the first sub-problem with lower complexity. Finally, the significance of the proposed approach over other baseline approaches is established through numerical analysis in terms of the \emph{MEAR} and fairness scores of the \emph{eMBB} UEs.  
\end{abstract}


%

\section{Introduction}

  The wireless industries are going through different kinds of emerging applications and services, e.g., high-resolution video streaming, virtual reality (VR), augmented reality (AR), autonomous cars, smart cities and factories, smart grids, remote medical diagnosis, unmanned aerial vehicles (UAV), artificial intelligence (AI) based personal assistants, sensing, metering, monitoring etc, along with the explosive trends of mobile traffic \cite{Cisco2017}.  It is foreseen that the mobile application market will flourish in a CAGR of $29.1\%$ during $2015-2020$ \cite{5GForum2016}. Energy efficiency, latency, reliability, data rate, etc are distinct for separate applications and services. To handle these diversified requirements, International Telecommunication Union (ITU) has already classified 5G services into \emph{uRLLC}, \emph{mMTC}, and \emph{eMBB} categories\cite{ITU2015}.  Gigabit per second (Gbps) level data rates are required for \emph{eMBB} users, whereas connection density and energy efficiency are the major concern for \emph{mMTC}, and \emph{uRLLC} traffic focuses on extremely high reliability ($99.999\%$) and remarkably low latency ($0.25\sim 0.30$ ms/packet)\cite{3GPP2016}.       
 
Generally, the lions' share of wireless traffic is produced by  \emph{eMBB} UEs. \emph{uRLLC} traffic is naturally infrequent and needs to be addressed spontaneously. The easiest way to settle this matter is to allocate some resources for \emph{uRLLC}. However, under-utilization of radio resources may emerge from this approach, and generally, effective multiplexing of traffics is required. For efficient multiplexing of  \emph{eMBB} and \emph{uRLLC} traffics, 3GPP has recommended a superposition/puncturing skeleton \cite{3GPP2016} and the short-TTI/puncturing approaches \cite{3GPP2017} in 5G cellular systems. Though the short-TTI mechanism is straightforward for implementation, it degrades spectral efficiency because of the massive overhead in the control channel. On the contrary, the puncturing strategy decreases the above overhead, although it necessitates an adequate mechanism for recognizing and healing the punctured case.  Slot ($1$ ms) and mini-slot ($0.125$ ms) are proposed as time units for meeting the latency requirement of \emph{uRLLC} traffic in the 5G NR. At the outset of a slot,  \emph{eMBB} traffic is scheduled and continues unchanged throughout the slot. If the same physical resources are used, \emph{uRLLC} traffic is overridden upon the scheduled  \emph{eMBB} transmission. 
 
 
 Currently, much attention has been paid to resource sharing for offering QoS or QoE to the users. Studies \cite{Bairagi2018Access} and \cite{Bairagi2018} investigate the sharing of an unlicensed spectrum between LTE and WiFi networks, however, the study \cite{Liu2017} con sider LTE-A and NB-IoT services for sharing the same resources. Study \cite{Abedin2018} solves user association and resource allocation problems. The study \cite{Abedin2018} consider the downlink of fog network to support QoS provisions of the \emph{uRLLC} and \emph{eMBB}. Some other studies, however, investigates and/or analyzes the influence of \emph{uRLLC} traffic on \emph{eMBB} \cite{Kassab2018,Ying2018,Pocovi2018,Popovski2018,Ji2018,Bennis2018} or presents architecture and/or framework for co-scheduling of \emph{eMBB} and \emph{uRLLC} traffic \cite{Liao2016,Pedersen2016,Pedersen2018K,Li2017}. Moreover, some authors consider \emph{eMBB} and \emph{uRLLC} traffic in their coexisting/multiplexing proposals  \cite{Wu2017,Anand2018,Pedersen2017,Bairagi2018KCC,Bairagi2019SAC,Pedersen2018,Esswie2018,Esswie2018July} where they apply puncturing technique.

As per our knowledge, concrete mathematical models and solutions, however, are lacking in most of these coexistence mechanisms. Most of the studies mainly focus on analysis, system-level design or framework. Thus, effective coexistence proposals between \emph{eMBB} and \emph{uRLLC} traffic are wanting in literature. So, to enable \emph{eMBB} and \emph{uRLLC} services in \emph{$5$G} wireless networks, we propose an effective coexistence mechanism in this paper. Our preliminary work has been published in \cite{Bairagi2019SAC} where we have used a one-sided matching and heuristic algorithm, respectively, for resolving resource allocation problems of \emph{eMBB} and \emph{uRLLC} users. The major difference between \cite{Bairagi2019SAC} and current work is the involvement of \emph{PSUM} and \emph{TM} for solving similar problems. This paper mainly focuses on the followings:
  \begin{itemize}
  	\item First, we formulate an optimization problem for \emph{eMBB} UEs with some constraints, where the objective is to maximize the minimum expected rate of \emph{eMBB} UEs over time. 
  	\item Second, to solve the optimization problem effectively, we decompose it into two sub-problems: resource scheduling for \emph{eMBB} UEs, and resource scheduling of \emph{uRLLC} UEs. \emph{PSUM} is used to solve the first sub-problem, whereas the \emph{TM} is employed to solve the second one.
  	\item Third, we redefine the first sub-problem into a minimization problem for each slot and provide an algorithm based upon \emph{PSUM} to obtain near-optimal solutions.
  	\item Fourth, we redefine the second sub-problem as a minimization problem for each mini-slot within every slot and present the algorithm based upon \emph{MCC} and \emph{MODI} methods of the transportation model to find an optimal solution of the second sub-problem.
  	\item Fifth, we also present a cost-effective heuristic algorithm for resolving the first sub-problem.       
  	\item Finally, we perform a comprehensive experimental analysis
  	for the proposed scheduling approach and compare the results, \emph{MEAR} and fairness\cite{Jain1984} of the \emph{eMBB} UEs, with the PS \cite{Pedersen2017}, MUPS\cite{Esswie2018}, RS, EDS, and MBS approaches.
  \end{itemize}
  
  The remainder of the paper is systematized as follows. In Section \ref{Lit_Rev}, we present the literature review. We explain the system model and present the problem formulation in Section \ref{Sys_model}. The proposed solution approach of the above-mentioned problem is addressed in Section \ref{Sol_dec}. In Section \ref{Per_eva}, we provide experimental investigation, discussion, and comparison concerning the proposed solution.  Finally, we conclude the paper in Section \ref{Conc}.
  
\section{Literature Review}
\label{Lit_Rev}
Recently, both industry and academia focus on the study of multiplexing between \emph{eMBB} traffic and \emph{uRLLC} traffic on the same physical resources. Information-theoretic arguments-based performance analysis for \emph{eMBB} and \emph{uRLLC}  traffic has performed in \cite{Kassab2018}. The authors consider both OMA and NOMA for uplink in C-RAN framework. An insight into the performance trade-offs among the \emph{eMBB} and \emph{uRLLC} traffic is explained in \cite{Kassab2018}.
In \cite{Ying2018}, authors have introduced \emph{eMBB} influenced minimization problem to protect the \emph{uRLLC} traffic from the dominant \emph{eMBB} services. This paper explores their proposal for the mobile front-haul environment.
In \cite{Pocovi2018}, the authors present an effective solution for multiplexing different traffics on a shared resource. Particularly, they propose an effective radio resource distribution method between the \emph{uRLLC} and \emph{eMBB} service classes following trade-offs among the reliability, latency and spectral efficiency. Moreover, they investigate the \emph{uRLLC} and \emph{eMBB} performance adopting different conditions.  

\begin{table}
	\centering
	\caption{List of Abbreviations}\label{Table_Abb}
	\renewcommand{\arraystretch}{1}
	\begin{tabular}{|c|p{12cm}|} \hline
		\textbf{Abbreviation} & \textbf{Elaboration}\\ \hline 
		uRLLC & Ultra-reliable Low-latency Communication\\ \hline
		eMBB, MBB & Enhanced Mobile Broadband, Mobile Broadband\\ \hline
		mMTC & Massive Machine-type Communication \\ \hline
		PSUM, SUM & Penalty Successive Upper bound Minimization, Successive Upper bound Minimization \\ \hline
		TTI & Transmission Time Interval\\ \hline
		NR & New Radio  \\ \hline
		QoS & Quality-of-Service  \\ \hline
		QoE & Quality-of-Experience  \\ \hline
		TM, BTM & Transportation Model, Balanced Transportation Model \\ \hline
		MCC &  Minimum Cell Cost \\ \hline
		MODI & Modified Distribution \\ \hline
		PS &  Punctured Scheduler \\ \hline
		MUPS & Multi-User Preemptive Scheduler \\ \hline
		RS &  Random Scheduler \\ \hline
		EDS & Equally Distributed Scheduler \\ \hline
		MBS & Matching Based Scheduler\\ \hline
		MEAR & Minimum Expected Achieved Rate \\ \hline
		NOMA, OMA &  Non-orthogonal Multiple Access, Orthogonal Multiple Access \\ \hline
		PRB, RB &  Physical Resource Block, Resource Block \\ \hline
		MIMO & Multiple-input Multiple-output \\ \hline
		SINR & signal-to-interference-noise-ratio\\ \hline
		gNB & Next Generation Base Station\\ \hline
		CP & Combinatorial Programming\\ \hline
		CDF, ECDF & Cumulative Distribution Function, Empirical Cumulative Distribution Function\\ \hline
		NWC & Northwest corner\\ \hline
		VAM & Vogel's Approximation Method\\ \hline
		MCS & Modulation and Coding Scheme\\ \hline
		CVaR & Conditional Value at Risk\\ \hline
		CAGR & Cumulative Average Growth Rate\\\hline
		C-RAN & Cloud Radio Access Network \\ \hline
	\end{tabular}
	\renewcommand{\arraystretch}{1}
	$\vspace{-.4cm}$
\end{table}

In order to 5G service provisioning (i.e., \emph{eMBB}, mMTC and \emph{uRLLC} services), the authors of \cite{Popovski2018} have studied radio resources slicing mechanism, where the performance of both orthogonal and non-orthogonal are analyzed. They have proposed a communication-theoretic model by considering the heterogeneity of 5G services. They also found that the non-orthogonal slicing is significantly better to perform instead of orthogonal slicing for those 5G service multiplexing.
Recently, for 5G NR physical layer challenges and solution mechanisms of \emph{uRLLC} traffic communications has been presented in \cite{Ji2018}, where they pay attention to the structure of packet and frame. Additionally, they focus on the improvement of scheduling and reliability mechanism for \emph{uRLLC} traffic communication such that the coexistence of \emph{uRLLC} with \emph{eMBB} is established.
In \cite{Bennis2018}, the authors have been analyzed the designing principle of the 5G wireless network by employing low-latency and high-reliability for \emph{uRLLC} traffic. To do this, they consider varying requirements of \emph{uRLLC} services such as variation of delay, packet size, and reliability. To an extent, they explore different topology network architecture under the uncertainty.

The authors of \cite{Liao2016, Pedersen2016, Pedersen2018K} present a resilient frame formation for multiplexing the provisions of different users. In \cite{Liao2016}, the authors jointly MBB and mission-critical communication  traffic by engaging dynamic TDD and TTI. In \cite{Pedersen2016}, the authors represent tractable multiplexing of MBB, \emph{MCC}, and mMTC considering dynamic TTI. The authors of \cite{Pedersen2018K} present a holistic overview of the agile scheduling for 5G that incorporates multiple users.  They envision an  E2E QoS architecture to offer improved opportunities for application-layer scheduling functionality that ensures QoE for each user. 
$M/D/m/m$ queueing model-based system-level design has proposed for fulfilling \emph{uRLLC} traffic demand in \cite{Li2017}, where they exhibit that the static bandwidth partitioning is inefficient for \emph{eMBB} and \emph{uRLLC} traffic. Thus, the authors of \cite{Li2017} have illustrated a dynamic mechanism for multiplexing of \emph{eMBB} and \emph{uRLLC} traffic and apply this in both frequency and time domain.

The efficient way of network resource sharing for the \emph{eMBB} and \emph{uRLLC} is studied in \cite{Wu2017} and \cite{Anand2018}.
A dynamic puncturing mechanism is proposed for \emph{uRLLC} traffic in \cite{Wu2017} within \emph{eMBB} resources to increase the overall resource utilization in the network.
To enhance the performance for decoding of \emph{eMBB} traffic, a joint signal space diversity and dynamic puncturing schemes have proposed, where they improve the performance of component interleaving as well as rotation modulation.
In \cite{Anand2018}, a joint scheduling problem is formulated for \emph{eMBB} and \emph{uRLLC} traffic in the goal of maximizing \emph{eMBB} users'utility while satisfying stochastic demand for the \emph{uRLLC} UEs. Specifically, they measure the loss of \emph{eMBB} users for superposition/puncturing by introducing three models, which include linear, convex and threshold-based schemes.
For reducing the queuing delay of the \emph{uRLLC} traffic, the authors introduce punctured scheduling (PS) in \cite{Pedersen2017}. In case of insufficient radio resource availability, the scheduler promptly overwrites a portion of the \emph{eMBB} transmission by the \emph{uRLLC} traffic. The scheduler improves the \emph{uRLLC} latency performance; however, the performance of the \emph{eMBB} users are profoundly deteriorated. The authors of \cite{Bairagi2018KCC} and \cite{Bairagi2019SAC} manifest the coexistence technique for enabling 5G wireless services like \emph{eMBB} and \emph{uRLLC} based upon a punctured scheme. The authors present an enhanced PS (EPS) scheduler to enable an improved ergodic capacity of the \emph{eMBB} users in \cite{Pedersen2018}.
EPS is capable of recovering the lost information due to puncturing and partially. \emph{eMBB} users are supposed to be cognizant about the corresponding resource that is being penetrated by \emph{uRLLC}. Therefore, the victim \emph{eMBB} users ignore the punctured resources from the erroneous chase condensing HARQ process. The authors of \cite{Esswie2018} propose a MUPS, where they discretize the trade-off among network system capacity and \emph{uRLLC} performance.
MUPS first tries to match the incoming \emph{uRLLC} traffic inside an \emph{eMBB} traffic in a conventional MU-MIMO transmission. MUPS serves the \emph{uRLLC} traffic instantly by using PS if multi-user (MU) pairing cannot be entertained immediately. Though MUPS shows improved spectral efficiency, it is not feasible for \emph{uRLLC} latency as  MU pairing mostly depends on the rate maximization. Hence, the inter-user interference can further degrade the SINR quality of the \emph{uRLLC} traffic, which can lead to reliability concerns. The authors of \cite{Esswie2018July} propose a null-space-based preemptive scheduler (NSBPS) for jointly serving \emph{uRLLC} and \emph{eMBB} traffic in a densely populated 5G arrangement. The proposed approach ensures on-the-spot scheduling for the sporadic \emph{uRLLC} traffic, while makes a minimal shock on the overall system outcome. The approach employs the system spatial degrees of freedom (SDoF) for \emph{uRLLC} traffic for spontaneously providing a noise-free subspace. In \cite{Alsenwi2019}, the authors present a risk-sensitive approach for allocating RBs to \emph{uRLLC} traffic in the goal of minimizing the uncertainty of \emph{eMBB} transmission. Particularly, they launch the Conditional Value at Risk (CVaR) for estimating the uncertainty of \emph{eMBB} traffic in \cite{Alsenwi2019}.

 \begin{figure}
 	\centering
 	\includegraphics[width=0.5\textwidth]{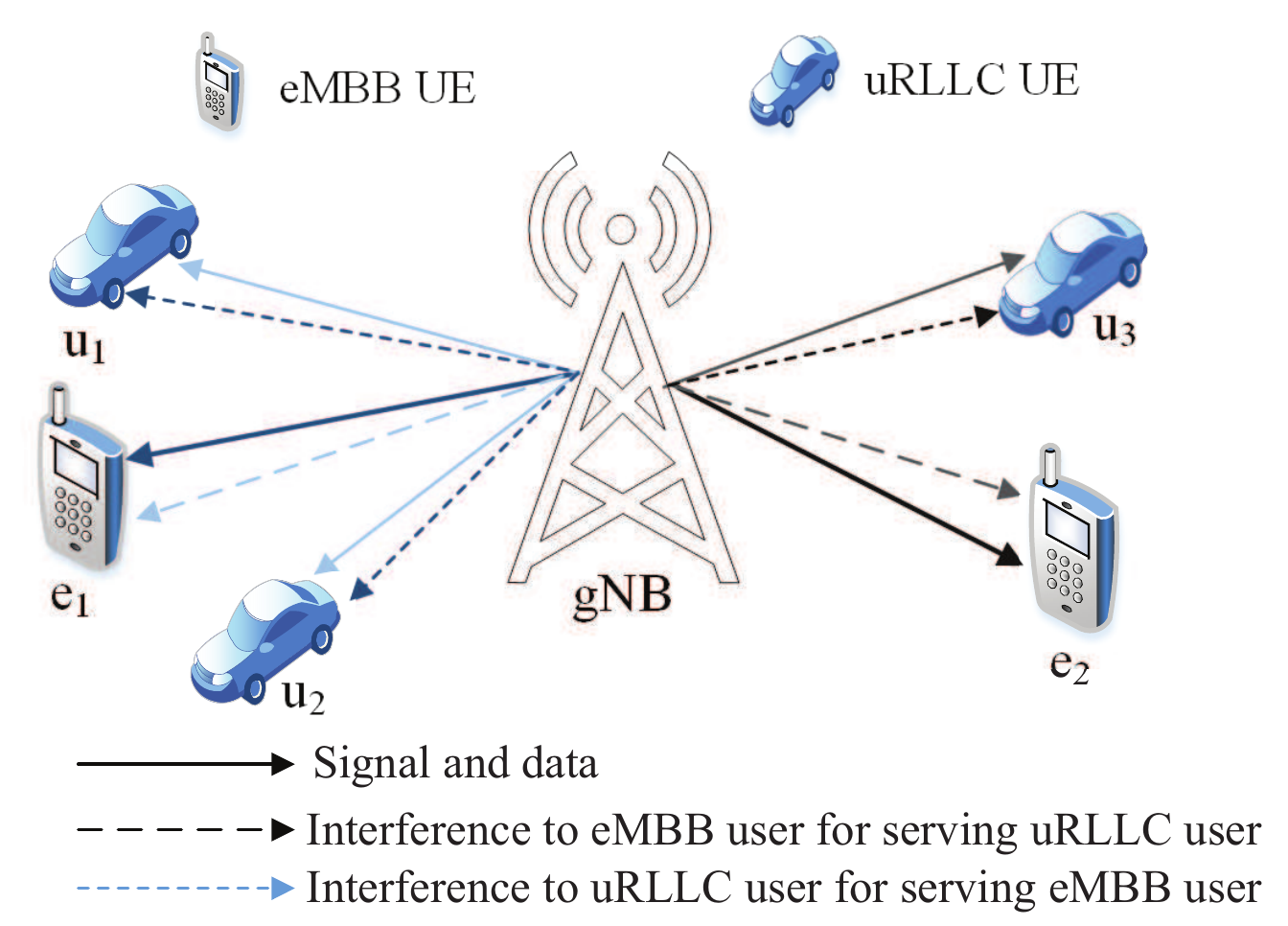}
 	\caption{System model for coexisting \emph{eMBB} and \emph{uRLLC} services in 5G.}\label{Fig1}
 \end{figure}

 \section{System Model and Problem Formulation}
 \label{Sys_model}
 In this work, we consider a 5G network scenario with one gNB which supports a group of user equipment (UE) $\mathcal{E}$ requiring \emph{eMBB} service, and a set of user equipment $\mathcal{U}$ demanding \emph{uRLLC} service. The system operates in downlink mode for the UEs and the overall system diagram is shown in Fig. \ref{Fig1}. gNB supports the UEs using licensed RBs $\mathcal{K}$ each with equal bandwidth of $B$. Every time slot, with a length $\Delta$, is split into $M$ mini-slots of duration $\delta$ for managing low latency services. For supporting \emph{eMBB} UEs, we consider $T_s$ LTE time slots  and denoted by $\mathcal{T}=\{1,2,\cdots, T_s\}$. \emph{uRLLC} traffic arrive at gNB (any mini-slot $m$ of time slot $t$) follows Gaussian distribution, i.e., $U\sim \mathcal{N}(\mu,\sigma^2)$. Here, $\mu$ and $\sigma^2$ denote the mean and variance of $U$. Each \emph{uRLLC} UE $u\in \mathcal{U}$ request for a payload of size $L_u^{m,t}$ (varying from 32 to 200 Bytes \cite{3GPPMar2017}).

  \begin{table}
  	\centering
  	\caption{Summary of Notations}\label{Table1}
  	\renewcommand{\arraystretch}{1}
  	\begin{tabular}{|c|p{12cm}|} \hline
  		\textbf{Symbol} & \textbf{Meaning}\\ \hline 
  		$\mathcal{E}$ & Set of active eMBB users\\ \hline
  		$\mathcal{U}$ & Set of uRLLC users \\ \hline
  		$\mathcal{K}$ & Set of RBs of uniform bandwidth $B$  \\ \hline
  		$B$ & Bandwidth of a RB  \\ \hline
  		$\Delta$ & Duration of a time slot  \\ \hline
  		$\delta$ & Duration of a mini-slot  \\ \hline
  		$M$ & Number of mini-slots in a time slot \\ \hline
  		$T$ & Total number of time slots \\ \hline
  		$\lambda$ & Mean value of arrival rate of uRLLC traffic \\ \hline
  		$U$ & \multicolumn{1}{p{10cm}|}{Random number representing arrival rate of traffics for uRLLC users at mini-slot $m$ of time slot $t$}\\ \hline
  		$L_u^{m,t}$ & \multicolumn{1}{p{10cm}|}{Payload size of uRLLC user $u\in \mathcal{U}$ at mini-slot $m$ of time slot $t$}\\ \hline
  		$\gamma_e^t$ & SNR of eMBB user $e\in \mathcal{E}$ in time slot $t$   \\ \hline
  		$P_e$ & Transmission power of gNB for eMBB user $e\in \mathcal{E}$\\ \hline
  		$h_e$ & Channel gain of for eMBB user $e\in \mathcal{E}$ from gNB\\ \hline
  		$N_0$ &  Noise spectral density\\ \hline
  		$\boldsymbol{\alpha}$ & \multicolumn{1}{p{10cm}|}{Resource allocation vector for $\mathcal{E}$}\\ \hline
  		$\gamma_u^{m,t}$ & \multicolumn{1}{p{10cm}|}{SINR/SNR of uRLLC user $u\in \mathcal{U}$ from gNB at mini-slot $m$ of time slot $t$}\\ \hline
  		$P_u$ & Transmission power of gNB for uRLLC user $u\in \mathcal{U}$\\ \hline
  		$h_u$ & Channel gain of for uRLLC user $u\in \mathcal{U}$ from gNB\\ \hline
  		$V_u$ & Channel dispersion for uRLLC user $u$ \\ \hline
  		$N_{u}^b$ & Blocklength of uRLLC traffic from user $u$ \\ \hline
  		$Q$ & Complementary Gaussian cumulative distribution function\\ \hline
  		$\varepsilon_u^d$ & Probability of decoding error for uRLLC user $u$ \\ \hline
  		$\boldsymbol{\beta}$ & \multicolumn{1}{p{10cm}|}{Resource allocation vector for $\mathcal{U}$}\\ \hline
  		$\boldsymbol{\phi}$ & \multicolumn{1}{p{6.5cm}|}{Vector for representing current serving uRLLC users}\\ \hline
  		$\epsilon$ & \multicolumn{1}{p{6.5cm}|}{uRLLC reliability probability}\\ \hline				
  		$r_{e,k}^t$ & \multicolumn{1}{p{10cm}|}{Achievable rate of eMBB user $e$ in RB $k$ of time slot $t$}\\ \hline
  		$r_{u,k}^{m,t}$ & \multicolumn{1}{p{10cm}|}{Achievable rate of uRLLC user $u$ in RB $k$ at mini-slot m of time slot $t$}\\ \hline
  	\end{tabular}
  	\renewcommand{\arraystretch}{1}
  	$\vspace{-.4cm}$
  \end{table}
  
  gNB allots the RBs to the \emph{eMBB} UEs at the commencement of any time slot $t\in \mathcal{T}$. The achievable rate of $e\in \mathcal{E}$ for RB $k\in \mathcal{K}$ is as follows:
  
 \begin{equation}
 \begin{split}
 r_{e,k}^t= \Delta B\log_2(1+\gamma_{e,k}^t),
 \end{split}
 \end{equation}
where $\gamma_{e,k}^t=\frac{P_eh_e^2}{N_0B}$ presents SNR. $P_e$ is the transmission power of gNB for $e\in \mathcal{E}$ and $h_e$ denotes the gain of  $e\in \mathcal{E}$ from the gNB, and $N_0$ represnts the noise spectral density. \emph{eMBB} UEs require more than one RB for satisfying their QoS. Therefore, the achievable rate of  \emph{eMBB} UE $e\in \mathcal{E}$ in time slot $t$ as follows:
 \begin{equation}
 \begin{split}
 r_{e}^t= \sum_{k\in \mathcal{K}} \alpha_{e,k}^t r_{e,k}^t,
 \end{split}
 \end{equation}
where $\boldsymbol{\alpha}$ denotes the resource allocation vector for $\mathcal{E}$ at any time slot $t$, and each element is as follows:
\begin{equation}
\alpha_{e,k}^t=
\begin{cases}
1, & \text{if RB}\  k \ \text{is allocated for $e \in \mathcal{E}$ at time slot $t$}, \\
0, & \text{otherwise}.
\end{cases}
\end{equation}

\emph{uRLLC} traffic can arrive at some moment (i.e. mini-slot) inside any time slot $t$ and requires to be attended quickly. Any \emph{uRLLC} traffic needs to be completed within a mini-slot period for its' latency and reliability constraints. Normally, the payload size of \emph{uRLLC} traffic is really short, and therefore, we cannot straightforwardly adopt Shannon's data rate formulation\cite{Kassab2018}. The achievable rate of a \emph{uRLLC} UE $u \in \mathcal{U}$ in RB $k\in \mathcal{K}$, when its' traffic is overlapped with \emph{eMBB} traffic,  can properly  be approximated by employing \cite{Scarlett2017} as follows:
\begin{equation}
\begin{split}
r_{u,k}^{m,t}= \delta\left[ B\log_2(1+\gamma_u^{m,t})-\sqrt{\frac{V_u}{N_u^b}}Q^{-1}(\varepsilon_u^d)\right],
\end{split}
\end{equation}

where $\gamma_u^{m,t}=\frac{h_u^2P_u}{N_0B+h_u^2P_e}$ represents the SINR for $u \in \mathcal{U}$ at mini-slot $m$ of  $t$. Here, $h_u^2P_e$ indicates the interference generated from serving $e\in \mathcal{E}$ in the same RB, $V_u=\frac{h_u^2P_u}{N_0B+h_u^2(P_u+P_e)}$ depicts the channel dispersion, and meaning of other symbols are shown in \ref{Table1}. However, the reliability of \emph{uRLLC} traffic fall into vulnerability due to the interference. Hence,  superposition mechanism is not a suitable for serving \emph{uRLLC} UE\cite{Ying2018}. Thus, for serving \emph{uRLLC} UEs, we concentrate on the puncturing technique . In the punctured mini-slot, gNB allots zero power for \emph{eMBB} UE, and therefore, the interference cannot affect the \emph{uRLLC} traffic. At that time, $\gamma_u^{m,t}=\frac{h_u^2P_u}{N_0B}$ and $V=\frac{h_u^2P_u}{N_0B+h_u^2P_u}$. The achieved rate of $u \in \mathcal{U}$, when it uses multiple RBs, is as follows:
\begin{equation}
\begin{split}
r_{u}^{m,t}= \sum_{k\in \mathcal{K}} \beta_{e,k}^{m,t} r_{u,k}^{m,t},
\end{split}
\end{equation}
where $\boldsymbol{\beta}$ is the resource allocation vector for $\mathcal{U}$ at $m$ of $t$, and each of its' element follows:
\begin{equation}
\beta_{e,k}^{m,t}=
\begin{cases}
1, & \text{if RB}\  k \ \text{is allocated for $u \in \mathcal{U}$ at $m$ of $t$}, \\
0, & \text{otherwise}.
\end{cases}
\end{equation}

\begin{figure}
	\centering
	\includegraphics[width=0.5\textwidth]{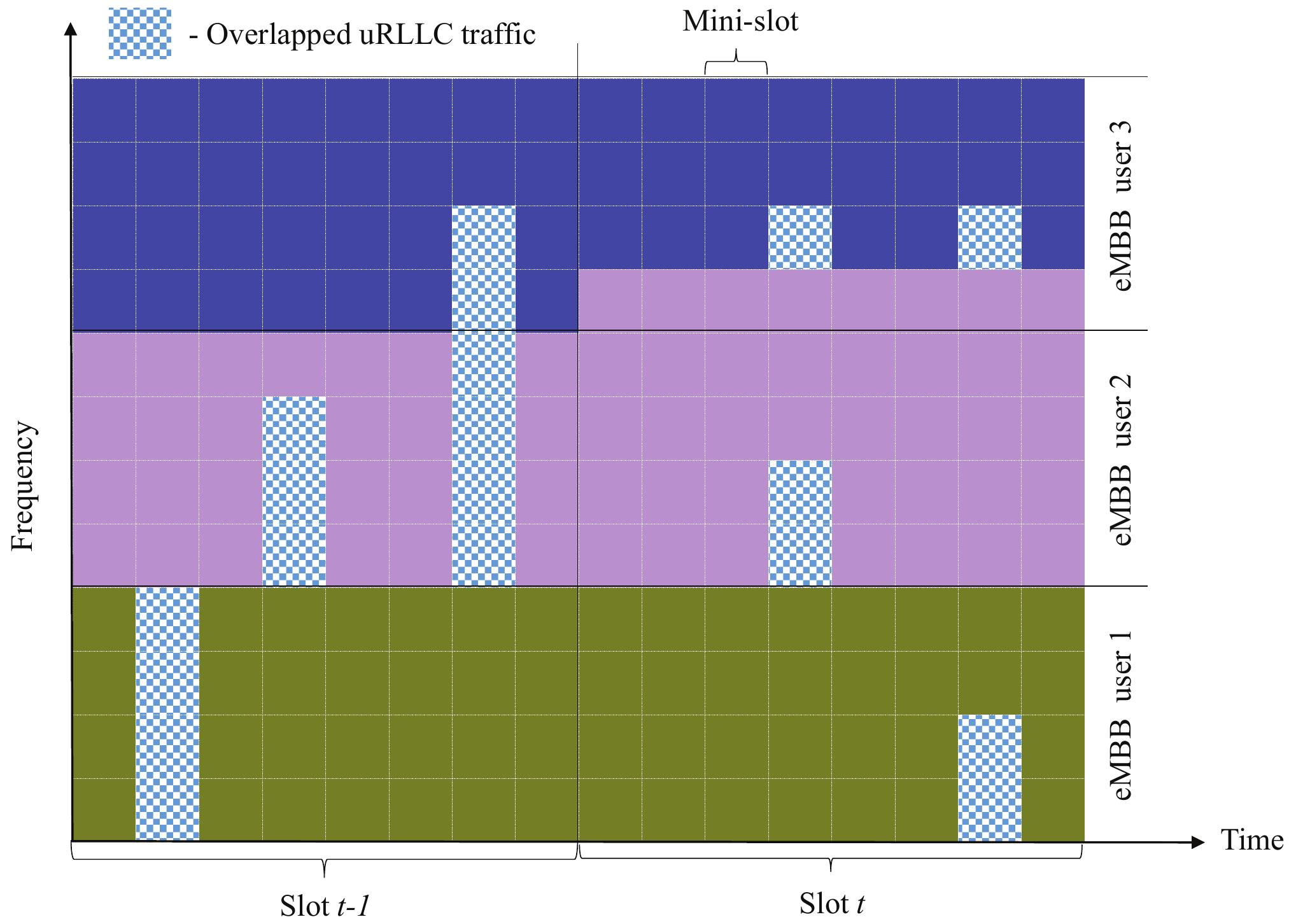}
	\caption{Example of multiplexing between \emph{eMBB} and \emph{uRLLC} traffic.}\label{Fig2}
\end{figure}

All the \emph{uRLLC} request in any $m$ of $t$ needs to be served for sure, and hence,  
\begin{equation}\label{Equ_Reliability}
\begin{split}
P(\sum_{u\in \mathcal{U}}\phi_u^{m,t}<U)\le \epsilon, \forall m,t.
\end{split}
\end{equation}
where $\boldsymbol{\phi}$ denotes a vector for the serving \emph{uRLLC} UEs, and thus, 
\begin{equation}
\phi_u^{m,t}=
\begin{cases}
1, & \text{if $u\in \mathcal{U}$ is served by the gNB at $m$ of $t$}, \\
0, & \text{otherwise}.
\end{cases}
\end{equation}

Within the stipulated period $\delta$, the payload $L_u^{m,t}$ of $u\in \mathcal{U}$ needs to be transferred, and hence, satisfy the following:
 \begin{equation}\label{Equ_Latency}
 \begin{split}
 \phi_u^{m,t}L_u^{m,t} \le \delta r_u^{m,t}, \forall u,m,t.
 \end{split}
 \end{equation}   

Hence, the reliability and latency concerns of \emph{uRLLC} traffic are simultaneously shielded by (\ref{Equ_Reliability}) and (\ref{Equ_Latency}). Besides, $e\in \mathcal{E}$ loses some throughput at $t$ if \emph{uRLLC} traffic is punctured within its' RBs. We utilize the linear model of \cite{Anand2018} for estimating the throughput-losses of \emph{eMBB} UE. Therefore, the throughput-losses $e\in \mathcal{E}$ looks like as follows:
\begin{equation}\label{Equ_loss}
\begin{split}
r_{e,loss}^{t}= \sum_{k\in \mathcal{K}}r_{e,k}^t\sum_{m\in \mathcal{M}}\sum_{u\in \mathcal{U}}\mathbb{I}(\alpha_{e,k}^t=\beta_{u,k}^{m,t}).
\end{split}
\end{equation}

So, the actual achievable rate of $e\in \mathcal{E}$ in any $t$ is as follows:
   \begin{equation}\label{Equ_actual}
   \begin{split}
   r_{e,actual}^t= r_e^t - r_{e,loss}^t.
   \end{split}
   \end{equation}

We see that $\boldsymbol{\beta}$ affects on $\boldsymbol{\alpha}$, and hence, impact negatively to the \emph{eMBB} throughput in each $t\in \mathcal{T}$. At the start of any  $t\in \mathcal{T}$, gNB allocates the RBs $\mathcal{K}$ among the $\mathcal{E}$ in an orthogonal fashion as shown in Fig. \ref{Fig2}. These characteristics of $\boldsymbol{\alpha}$ are shown mathematically as follows: 
 \begin{equation} \label{Equ_alpha1}
 \begin{split}
 \sum_{e\in \mathcal{E}} \alpha_{e,k}^t &\le 1, \forall k,\\
 \end{split}
 \end{equation}
 \begin{equation} \label{Equ_alpha2}
 \begin{split}
 \sum_{k\in \mathcal{K}} \alpha_{e,k}^t &\ge 1, \forall e,\\
 \end{split}
 \end{equation}
 \begin{equation} \label{Equ_alpha3}
 \begin{split}
 \sum_{e\in \mathcal{E}}\sum_{k\in \mathcal{K}} \alpha_{e,k}^t &\le |\mathcal{K}|.\\
 \end{split}
 \end{equation}
 
 Within each $t\in \mathcal{T}$, gNB allows \emph{uRLLC} UEs to get some RBs immediately on a mini-slot basis. Therefore, \emph{uRLLC} traffic overlaps with \emph{eMBB} traffic at $m$ and also shown in Fig. \ref{Fig2}.  Accordingly, $\boldsymbol{\beta}$ satisfy the following conditions on each $m$:

\begin{equation}\label{Equ_beta1}
\begin{split}
\sum_{u\in \mathcal{U}} \beta_{u,k}^{m,t} &\le 1, \forall k,
\end{split}
\end{equation}
\begin{equation}\label{Equ_beta2}
\begin{split}
\sum_{k\in \mathcal{K}} \phi_u^{m,t} \beta_{u,k}^{m,t} &\ge 1, \forall u,
\end{split}
\end{equation}
\begin{equation}\label{Equ_beta3}
\begin{split}
\sum_{u\in \mathcal{U}}\sum_{k\in \mathcal{K}} \phi_u^{m,t} \beta_{u,k}^{m,t} &\le |\mathcal{K}|.
\end{split}
\end{equation} 

Finally, our objective is to maximize the actual achievable rate of each \emph{eMBB} UE across $\mathcal{T}$ while entertaining nearly every \emph{uRLLC} request within its' speculated latency. We apply $\emph{Max-Min}$  fairness doctrine for this mission, and it contributes stationary service quality, enhances spectral efficiency and makes UEs more pleasant in the network.  Hence, the maximization problem is formulated as follows:

\begin{subequations}\label{Opt1}
	\begin{align}
	\underset{\boldsymbol{\alpha},\boldsymbol{\beta}} \max \
	& \underset{e\in \mathcal{E}}\min \ \mathbb{E}\left(\sum_{t=1}^{|\mathcal{T}|}r_{e,actual}^t \right)   \tag{\ref{Opt1}} \\
	\text{s.t.} \quad &\label{Opt1:const1} P\left(\sum_{u\in \mathcal{U}}\phi_u^{m,t}< U\right)\le \epsilon, \forall m,t, \\ 
	&\label{Opt1:const2} \phi_u^{m,t}L_u^{m,t} \le \delta r_u^{m,t}, \forall u,m,t,\\
	&\label{Opt1:const3} \sum_{e\in \mathcal{E}} \alpha_{e,k}^t \le 1, \forall k,t,\\
	&\label{Opt1:const4} \sum_{u\in \mathcal{U}} \beta_{u,k}^{m,t} \le 1, \forall k,m,t,\\
	&\label{Opt1:const5} \sum_{k\in \mathcal{K}} \alpha_{e,k}^t \ge 1, \forall e,t,\\
	&\label{Opt1:const6} \sum_{k\in \mathcal{K}} \phi_u^{m,t} \beta_{u,k}^{m,t} \ge 1, \forall u,m,t,\\
	&\label{Opt1:const7} \sum_{e\in \mathcal{E}}\sum_{k\in \mathcal{K}} \alpha_{e,k}^t + \sum_{u\in \mathcal{U}}\sum_{k\in \mathcal{K}} \phi_u^{m,t} \beta_{u,k}^{m,t}\le |\mathcal{K}|, \forall t, \\
	&\label{Opt1:const9} \alpha_{e,k}^{t},\beta_{u,k}^{m,t},\phi_u^{m,t}\in \{0,1\}, \forall e,u,k,m,t.
	\end{align}
\end{subequations}
In (\ref{Opt1}), the reliability and latency constraints of the \emph{uRLLC} UEs are preserved by (\ref{Opt1:const1}) and (\ref{Opt1:const2}). Constraints (\ref{Opt1:const3}) and (\ref{Opt1:const4}) are used to show the orthogonality of RBs among \emph{eMBB} and \emph{uRLLC} UEs, respectively. At least one RB is posed by every active UE and is encapsulated by both (\ref{Opt1:const5}) and (\ref{Opt1:const6}). Resource restriction is presented by constraint (\ref{Opt1:const7}). Constraint (\ref{Opt1:const9}) shows that  every item of $\boldsymbol{\alpha}$, $\boldsymbol{\beta}$ and $\boldsymbol{\phi}$ are binary. The formulation (\ref{Opt1}) is a Combinatorial Programming (CP) problem having chance constraint, and NP-hard due to its nature.


\begin{figure}
	\centering
	\includegraphics[width=0.5\textwidth]{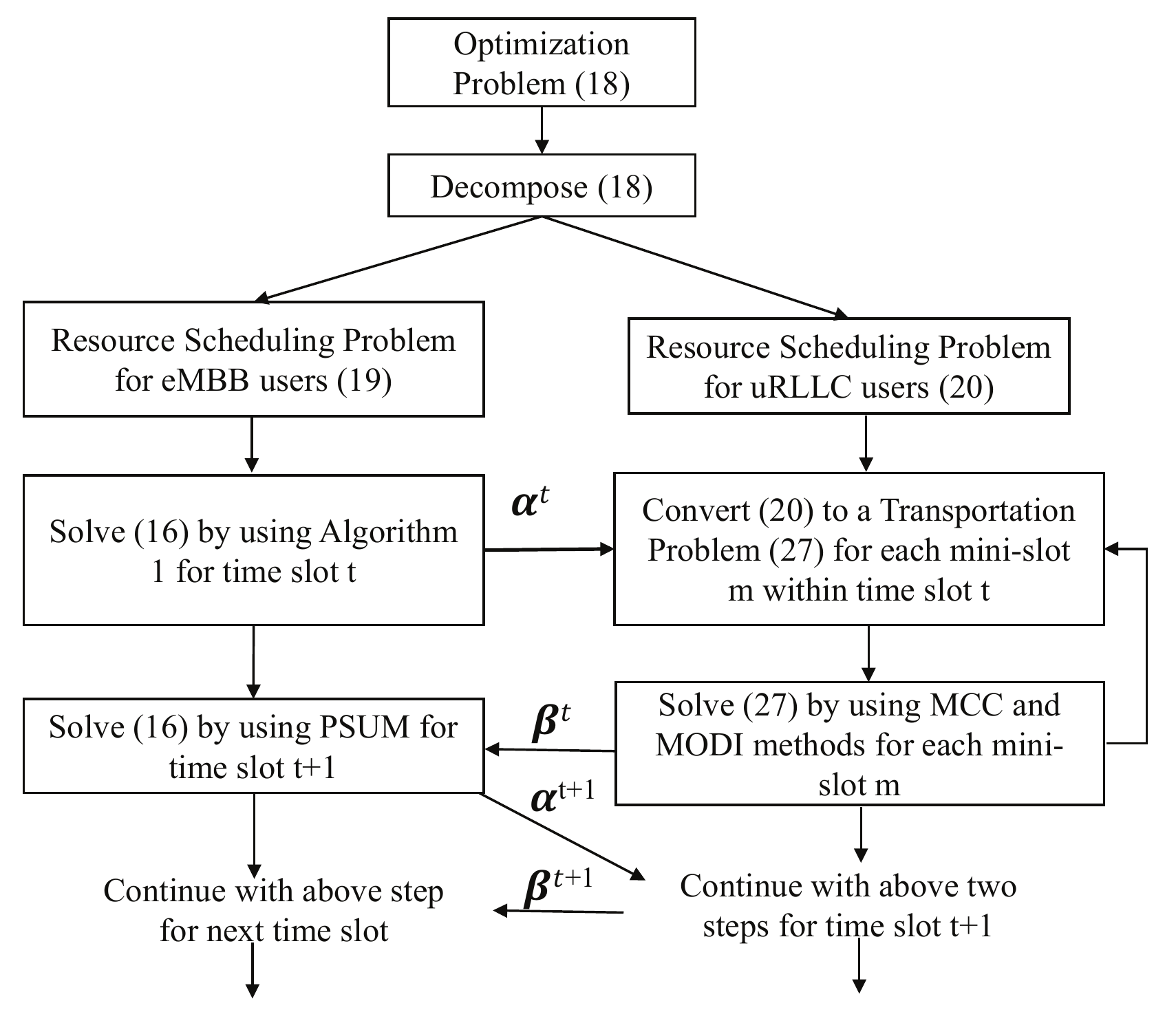}
	\caption{Overview of the Solution Process for (\ref{Opt1}).}\label{Fig_Overall_sol}
\end{figure}

\section{Decomposition as a Solution Approach for problem (\ref{Opt1})}
\label{Sol_dec}
We assume that \emph{eMBB} UEs are data-hungry over the considered period. Thus, at the commencement of a time slot $t\in \mathcal{T}$, gNB schedules all of its' RBs among the \emph{eMBB} UEs and stay unchanged over $t$. If \emph{uRLLC} traffic requests come in any $m$ of $t$, the scheduler tries to serve the requests in the next $m+1$. Hence, the overlapping of \emph{uRLLC} traffic over \emph{eMBB} traffic happens as shown in Fig. \ref{Fig2}. Usually, a portion of all RBs is required for serving such \emph{uRLLC} traffic. However, the challenge is to find the victimized \emph{eMBB} UE(s) following the aspiration of the problem (\ref{Opt1}). 

For getting an effective solution to the problem (\ref{Opt1}), we can utilize the concept of a divide-and-conquer strategy. Here, we divide (\ref{Opt1}) into two resource allocation sub-problems, namely,  for \emph{eMBB} UEs on time slot basis and \emph{uRLLC} UEs on a mini-slot basis. The first sub-problem is as follows:
\begin{subequations}\label{Opt1_1}
	\begin{align}
	\underset{\boldsymbol{\alpha}} \max \
	& \underset{e\in \mathcal{E}}\min\ \mathbb{E}\left(\sum_{t=1}^{|\mathcal{T}|}r_{e,actual}^t \right)  \tag{\ref{Opt1_1}} \\
	\text{s.t.} \quad & \label{Opt1_1:const1} \sum_{e\in \mathcal{E}} \alpha_{e,k}^t \le 1, \forall k,t,\\
	&\label{Opt1_1:const2} \sum_{k\in \mathcal{K}} \alpha_{e,k}^t \ge 1, \forall e,t,\\
	&\label{Opt1_1:const3} \sum_{e\in \mathcal{E}}\sum_{k\in \mathcal{K}} \alpha_{e,k}^t \le |\mathcal{K}|, \forall t, \\
	&\label{Opt1_1:const4} \alpha_{e,k}^{t}\in \{0,1\}, \forall e,k,t.
	\end{align}
\end{subequations}

On the other hand, the second sub-problem (with $\alpha^t,\forall t$ as the solution of \ref{Opt1_1}) is manifested as follows:

\begin{subequations}\label{Opt1_2}
	\begin{align}
	\underset{\boldsymbol{\beta}} \max \
	& \underset{e\in \mathcal{E}}\min \ \mathbb{E}\left(\sum_{t=1}^{|\mathcal{T}|}r_{e,actual}^t\right)   \tag{\ref{Opt1_2}} \\
	\text{s.t.} \quad &\label{Opt1_2:const1} P\left(\sum_{u\in \mathcal{U}}\phi_u^{m,t}< U \right)\le \epsilon, \forall m,t, \\ 
	&\label{Opt1_2:const2} \phi_u^{m,t}L_u^{m,t} \le \delta r_u^{m,t}, \forall u,m,t,\\
	&\label{Opt1_2:const3} \sum_{u\in \mathcal{U}} \beta_{u,k}^{m,t} \le 1, \forall k,m,t,\\
	&\label{Opt1_2:const4} \sum_{k\in \mathcal{K}} \phi_u^{m,t} \beta_{u,k}^{m,t} \ge 1, \forall u,m,t,\\
	&\label{Opt1_2:const5} \sum_{u\in \mathcal{U}}\sum_{k\in \mathcal{K}} \phi_u^{m,t} \beta_{u,k}^{m,t} \le |\mathcal{K}|, \forall m,t,\\
	&\label{Opt1_2:const6} \beta_{u,k}^{m,t},\phi_u^{m,t}\in \{0,1\}, \forall u,k,m,t.
	\end{align}
\end{subequations}

Fig. \ref{Fig_Overall_sol} shows the solution overview of the optimization problem (\ref{Opt1}). We can better understand the philosophy of the problem and the solution approach with an illustrative example in Fig. \ref{Fig2}. At the beginning of the time slot, $t-1$, let us assume that there are 3 \emph{eMBB} UEs, each of whom owns 4 RBs. Within $t-1$, the service request for \emph{uRLLC} UEs came abruptly and the allocation of RBs for that UEs is shown in Fig. \ref{Fig2}, as overlapped \emph{uRLLC} traffic in the mini-slots. During this time, \emph{eMBB} users $1,2$ and $3$ waste throughput equivalent to $4$RBs$\times 1$ mini-slot, $7$RBs$\times 1$ mini-slot, and $2$RBs$\times 1$ mini-slot, respectively. At the start of the next time slot, $t$, gNB acknowledges the resource scheduling of \emph{uRLLC} UEs of $t-1$ to allocate and compensate \emph{eMBB} UEs. gNB allocates more RBs to \emph{eMBB} user 2 and less to \emph{eMBB} user 3 as they lose more and less, respectively, in the time slot $t-1$. Moreover, EgNB tries to serve \emph{uRLLC} users such that the loss of throughput of \emph{eMBB} users are almost similar in the time slot $t$. Therefore, gNB makes a balance among the throughput of \emph{eMBB} users in each time slot, which ultimately serves to reach the goal of (\ref{Opt1}) on a long-run basis.  

\subsection{\emph{PSUM} as a Solution of the Sub-Problem (\ref{Opt1_1})}
\label{Sol_Opt1_1}
Problem (\ref{Opt1_1}) is still is computationally expensive to reach a globally optimal solution due to its' NP-hardness. In this sub-section, we propose the \emph{PSUM} algorithm to solve (\ref{Opt1_1}) approximately with low complexity.  Relaxation of the binary variable and the addition of a penalty term to the objective function is the main philosophy of our proposed \emph{PSUM} algorithm. We redefine  (\ref{Opt1_1}) as follows:
\begin{subequations}\label{Opt1_1_2}
	\begin{align}
	\underset{\boldsymbol{\alpha^t}} \min
	& \sum_{e\in \mathcal{E}}W_e^t(\boldsymbol{\alpha^t}),\forall t,   \tag{\ref{Opt1_1_2}} \\
	\text{s.t.} \quad & \label{Opt1_1_2:const3}(\ref{Opt1_1:const1}),(\ref{Opt1_1:const2}),(\ref{Opt1_1:const3}),\\
	&\label{Opt1_1_2:const4} W_e^t(\boldsymbol{\alpha^t})=\bigg|\frac{1}{t|\mathcal{E}|}\sum_{e'\in \mathcal{E}}\bigg(\sum_{t'=1}^{t-1}r_{e',actual}^{t'}+r_{e}^{t}\bigg) \notag\\
	& \qquad \qquad -\frac{1}{t}\bigg(\sum_{t'=1}^{t-1}r_{e,actual}^{t'}+r_{e}^{t}\bigg)\bigg|,\\
	&\label{Opt1_1_2:const5} \alpha_{e,k}^{t}\in [0,1], \forall e,k,t.
	\end{align}
\end{subequations}

Now according to Theorem 2 of \cite{Zhang2017}, if $|\mathcal{K}|$ is sufficiently large then original sub-problem (\ref{Opt1_1}) and (\ref{Opt1_1_2}) are equivalent. Moreover, we add a penalty term $L_p$ to the objective function to get binary soltion of relaxed variable from (\ref{Opt1_1_2}). Let $\boldsymbol{\alpha}_k^t=\{\alpha_{e,k}^t\}_{e\in \mathcal{E}}$ and we can rewrite (\ref{Opt1_1:const1}) as $\parallel\boldsymbol{\alpha}_k^t\parallel_1\le 1,\forall t,k$. The penalized problem is as follows:
  \begin{subequations}\label{Opt1_1_3}
  	\begin{align}
  	\underset{\boldsymbol{\alpha^t}} \min
  	& \sum_{e\in \mathcal{E}}W_e^t(\boldsymbol{\alpha^t})+\sigma P_{\varepsilon}(\boldsymbol{\alpha^t}),\forall t   \tag{\ref{Opt1_1_3}} \\
  	\text{s.t.} \quad & (\ref{Opt1_1_2:const3}),(\ref{Opt1_1_2:const4}),(\ref{Opt1_1_2:const5}),
  	\end{align}
  \end{subequations}
where $\sigma>0$ is the penalty parameter,
\begin{equation}
\begin{split}
P_{\varepsilon}(\boldsymbol{\alpha^t})=\sum_{k\in \mathcal{K}}(\parallel \boldsymbol{\alpha}_k^t + \varepsilon\boldsymbol{1}\parallel_p^p-c_{\varepsilon,k}).
\end{split}
\end{equation}
 with $p\in (0,1)$, and $\varepsilon$ is any non-negative constant. Following the fact of \cite{Liu2015} which is further described in \cite{Zhang2017}, the optimal value is as follows:
 \begin{equation}
 \begin{split}
c_{\varepsilon,k}=(1+\varepsilon)^p+(|\mathcal{E}|-1)\varepsilon^p.
 \end{split}
 \end{equation}
 Generally, the parameter $\sigma$ should big enough to make the values of $\{\alpha_{e,k}^t\}$ near zero or one. Then, we achieve a feasible solution of (\ref{Opt1_1_3}) by applying the rounding process.

\begin{algorithm}
	\caption{Solution of (\ref{Opt1_1}) for each $t$ based on \emph{PSUM}}\label{Algo1}
	\begin{algorithmic}[1]
		\STATE $\mathbf{Initialization}$: $\varepsilon_1,\sigma_1,I_{max}$ and let $i=0$ 
		\STATE Solve problem (\ref{Opt1_1_2}) and obtain solution $\boldsymbol{\alpha}^{t,0}$
		\WHILE {$i<I_{max}$} 
		\STATE Set $\varepsilon=\varepsilon_{i+1}$ and $\sigma=\sigma_{i+1}$
		\STATE Solve problem (\ref{Opt1_1_4}) with the initial point being $\boldsymbol{\alpha}^{t,i}$, and obtain a new solution $\boldsymbol{\alpha}^{t,i+1}$
		\IF {$\boldsymbol{\alpha}^{t,i+1}$ is binary}
		\STATE Stop
		\ELSE
		\STATE Set $i=i+1$ 
		\STATE Update $\varepsilon_{i+1}=\eta \varepsilon$, and $\sigma_{i+1}=\zeta \sigma$
		\ENDIF   
		\ENDWHILE
	\end{algorithmic} 
\end{algorithm}

It is not easy to solve (\ref{Opt1_1_3}) directly. However, by utilizing the successive upper bound minimization (SUM) technique \cite{Hunter2004, Razaviyayn2013}, we can efficiently resolve (\ref{Opt1_1_3}). This method tries to secure the lower bound of the actual objective function by determining a sequence of approximation of the objective functions. As $P_{\varepsilon}(\boldsymbol{\alpha}^t)$ is concave in nature and hence, 
\begin{equation}
\begin{split}
P_{\varepsilon}(\boldsymbol{\alpha}^t)\le P_{\varepsilon}(\boldsymbol{\alpha}^{t,i})+\nabla P_{\varepsilon}(\boldsymbol{\alpha}^{t,i})^T(\boldsymbol{\alpha}^t-\boldsymbol{\alpha}^{t,i}),
\end{split}
\end{equation}
where $\boldsymbol{\alpha}^{t,i}$ is the value of current allocation of iteration $i$. At the $(i+1)$-th iteration of $t$, we solve the following problem:
      \begin{subequations}\label{Opt1_1_4}
      	\begin{align}
      	\underset{\boldsymbol{\alpha^t}} \min
      	& \sum_{e\in \mathcal{E}}W_e^t(\boldsymbol{\alpha^t})+\sigma_{i+1} \nabla P_{\varepsilon}(\boldsymbol{\alpha}^{t,i})^T \boldsymbol{\alpha^{t}}   \tag{\ref{Opt1_1_4}} \\
      	\text{s.t.} \quad & (\ref{Opt1_1_2:const3}),(\ref{Opt1_1_2:const4}),(\ref{Opt1_1_2:const5}).
      	\end{align}
      \end{subequations} 

In each iteration, we can get a globally optimal solution for sub-problem (\ref{Opt1_1_4}) by using the solver. Algorithm \ref{Algo1} shows the proposed mechanism for solving (\ref{Opt1_1}). In this Algorithm, $0<\eta<1<\zeta$ where $\zeta$ and $\eta$ represent two constants defined previously.

\subsection{Solution of Sub-Problem (\ref{Opt1_2}) through TM}
\label{Sol_Opt1_2}
Due to the existence of chance constraint (\ref{Opt1_2:const1}) and also the combinatorial variable, $\boldsymbol{\beta}$, (\ref{Opt1_2}) is still difficult to resolve by using traditional optimizer. Now, we need to transmute (\ref{Opt1_2:const1}) into deterministic form for solving  (\ref{Opt1_2}). Moreover, let us assume $g(\boldsymbol{\phi},U)=\sum_{u\in \mathcal{U}}\phi_u^{m,t}-U$, $U\in \mathbb{R}$ and $U\sim \mathcal{N}(\mu,\sigma^2),\forall m,t$ and hence,
 \begin{subequations}\label{Chance_1}
\begin{align}
 Pr\{g(\phi,U)\le 0\}=& Pr\bigg\{\sum_{u\in \mathcal{U}}\phi_u^{m,t}-U\le 0\bigg\} \tag{\ref{Chance_1}}\\
                     =& \label{Chance_1:line2} Pr\bigg\{\sum_{u\in \mathcal{U}}\phi_u^{m,t}\le U\bigg\}\\
                     =& \label{Chance_1:line3}1-Pr\bigg\{\sum_{u\in \mathcal{U}}\phi_u^{m,t}\ge U\bigg\}\\
                     =& \label{Chance_1:line4}1-Pr\bigg\{\frac{U-\mu}{\sigma}\le \frac{\sum_{u\in \mathcal{U}}\phi_u^{m,t}-\mu}{\sigma} \bigg\}\\
                     =& \label{Chance_1:line5} 1-F_{U}\left(\sum_{u\in \mathcal{U}}\phi_u^{m,t}\right).
\end{align}
\end{subequations} 

Here, $F_U$ is the cumulative distribution function (CDF) of random variable $U$. Thus, from constraint (\ref{Opt1_2:const1}), we can rewrite as follows:
  \begin{subequations}\label{Chance_2}
   \begin{align}
   &Pr\{g(\phi,U)\le 0\}\ge \epsilon \tag{\ref{Chance_2}},\\
  & \label{Chance_2:line2}1-F_{U}\bigg(\sum_{u\in \mathcal{U}}\phi_u^{m,t}\bigg) \le \epsilon, \\
  & \label{Chance_2:line3}F_{U}\bigg(\sum_{u\in \mathcal{U}}\phi_u^{m,t}\bigg) \ge 1-\epsilon ,\\
  & \label{Chance_2:line4}\sum_{u\in \mathcal{U}}\phi_u^{m,t} \ge F_{U}^{-1} \bigg(1-\epsilon\bigg) ,\\
  & \label{Chance_2:line5}\sum_{u\in \mathcal{U}}\phi_u^{m,t} -F_{U}^{-1} (1-\epsilon) \ge 0.  
   \end{align}
\end{subequations} 

Now, (\ref{Chance_2:line5}) and (\ref{Opt1_2:const1}) are identical. Hence, the renewed form of (\ref{Opt1_2}) looks like as follows:
 \begin{subequations}\label{Opt1_2_1}
 	\begin{align}
 	\underset{\boldsymbol{\beta}^t} \min
 	\quad & \sum_{e\in \mathcal{E}}V_e^t(\boldsymbol{\alpha^t},\boldsymbol{\beta}^t),\forall t \tag{\ref{Opt1_2_1}} \\
 	\text{s.t.} \quad &\label{Opt1_2_1:const1} \sum_{u\in \mathcal{U}}\phi_u^{m,t}-F_{U}^{-1} (1-\epsilon) \ge 0, \forall m, \\ 
 	\label{Opt1_2_1:const2} \quad & (\ref{Opt1_2:const2}),(\ref{Opt1_2:const3}),(\ref{Opt1_2:const4}),(\ref{Opt1_2:const5}),(\ref{Opt1_2:const6}), \forall u,m,\\
 	\label{Opt1_2_1:const3}\quad & V_e^t(\boldsymbol{\alpha^t},\boldsymbol{\beta}^t)=\bigg|\frac{1}{|\mathcal{E}|}\sum_{e^{'}\in \mathcal{E}}r_{e^{'},loss}^t-r_{e,loss}^t\bigg|, \forall e.
 	\end{align}
 \end{subequations}
 
 Problem (\ref{Opt1_2_1}) is still NP-hard due to the appearance of combinatorial variable. In (\ref{Opt1_2_1}), (\ref{Opt1_2_1:const1}) holds for a particular value of  $\epsilon$ when gNB serves a certain portion of \emph{uRLLC} UE  $U^{'}\le U$. For a $m$ of $t$, let us assume $\mathcal{U}^{'}=\{1,2,....,U^{'}\}$ and $\phi_u^{m,t}=1,\forall u\in \mathcal{U}^{'}$. We can determine the requisite RBs, $\forall u\in \mathcal{U}^{'}$ holding $\delta$ as the upper-bound in (\ref{Opt1_2:const2}) and let $\boldsymbol{d}=[d_1,d_2,\cdots,d_{|\mathcal{U}^{'}|}]$. As gNB engages OFDMA for \emph{uRLLC} UEs, constraint (\ref{Opt1_2:const3}) holds. Moreover, depending on $\mathcal{U}^{'}$, constraints (\ref{Opt1_2:const4}), (\ref{Opt1_2:const5}), and (\ref{Opt1_2:const6}) also hold. Constraint (\ref{Opt1_2_1:const3}) can be used as a basic block to build a cost matrix $\boldsymbol{C}=(c_{u,e}), u\in \mathcal{U}^{'},e\in \mathcal{E}$. As $\mathcal{K}$ are held by \emph{eMBB} UEs $\mathcal{E}$ in any time slot $t\in \mathcal{T}$, we can find a vector $\boldsymbol{s}=[s_1,s_2,\cdots,s_{|\mathcal{E}|}]$. Now redefine problem (\ref{Opt1_2_1}) as follows:
 
\begin{subequations}\label{Opt1_2_2}
	\begin{align}
	\underset{\boldsymbol{\chi}} \min
	\quad & \sum_{u\in \mathcal{U}^{'}}\sum_{e\in \mathcal{E}}c_{ue}\chi_{ue} \tag{\ref{Opt1_2_2}} \\
	\text{s.t.} \quad &\label{Opt1_2_2:const1} \sum_{e\in \mathcal{E}} \chi_{ue}=d_u,\forall u\in \mathcal{U}^{'}, \\ 
	\label{Opt1_2_2:const2} \quad & \sum_{u\in \mathcal{U}^{'}}\chi_{ue}\le s_e, \forall e\in \mathcal{E},\\
	\label{Opt1_2_2:const3}\quad & \sum_{u\in \mathcal{U}^{'}}d_u \le \sum_{e\in \mathcal{E}} s_e,\\
	\label{Opt1_2_2:const4}\quad & \sum_{e\in \mathcal{E}} s_e = |\mathcal{K}|,\\
	\label{Opt1_2_2:const5}\quad & \chi_{ue}\ge 0,\forall u\in \mathcal{U}^{'},e\in \mathcal{E}.
	\end{align}
\end{subequations}
The goal of (\ref{Opt1_2_2}) is to find a matrix $\boldsymbol{\chi}\in \mathbb{Z}^{|\mathcal{U}^{'}|\times|\mathcal{E}|}=(\chi_{ue}), \forall u\in \mathcal{U}^{'},e\in \mathcal{E}$ that will minimize the cost/loss of \emph{eMBB} UEs. This is a linear programming problem equivalent to the Hitchcock problem \cite{Hitchcock1941} with inequities, which contributed to unbalanced transportation model. Introducing slack variables $\chi_{|\mathcal{U}^{'}|+1,e}, \forall e\in \mathcal{E}$ and $d_{|\mathcal{U}^{'}|+1}$ in the constraints (\ref{Opt1_2_2:const2}) and (\ref{Opt1_2_2:const3}), respectively, which convert them into equality, we have:
 \begin{equation}\label{Equ_Const24b}
 \begin{split}
 \sum_{u\in \mathcal{U}^{'}}\chi_{ue}+\chi_{|\mathcal{U}^{'}|+1,e} = s_e, \forall e\in \mathcal{E},
 \end{split}
 \end{equation}
 
 \begin{equation}\label{Equ_Const24c}
 \begin{split}
 \sum_{u\in \mathcal{U}^{'}}d_u + d_{|\mathcal{U}^{'}|+1} = \sum_{e\in \mathcal{E}} s_e.
 \end{split}
 \end{equation}

Now the modified problem in (\ref{Opt1_2_2}) is a \emph{BTM}. Moreover, we have to add $d_{|\mathcal{U}^{'}|+1}=\sum_{e\in \mathcal{E}} s_e-\sum_{u\in \mathcal{U}^{'}}d_u$ to the demand vector $\boldsymbol{d}$ as $\boldsymbol{d}=\boldsymbol{d} \cup \{d_{|\mathcal{U}^{'}|+1}\}$ and a row $[0]_{1\times |\mathcal{E}|}$ to cost matrix $\boldsymbol{C}$ as $\boldsymbol{C}=\boldsymbol{C}\cup \{[0]_{1\times |\mathcal{E}|}\}$. \emph{BTM} can be solved by the simplex method \cite{Dantzig1963}. The solution matrix $\boldsymbol{\chi}$ will be in the form of $\mathbb{Z}^{(|\mathcal{U}^{'}|+1) \times|\mathcal{E}|}$. \emph{NWC}\cite{Reinfeld1958}, \emph{MCC}\cite{Reinfeld1958}, and VAM\cite{Reinfeld1958,Shimshak1981} are some of the popular methods for obtaining initial feasible solution of \emph{BTM}. We can use the stepping-stone \cite{Charnes1954} or \emph{MODI} \cite{Taha2007} method to get an optimal solution of the \emph{BTM}. In the following sub-section, we use the combination of the \emph{MCC} and \emph{MODI} for acquaring the optimal result from the \emph{BTM}.

\subsubsection{Determining Initial Feasible Solution by \emph{MCC} Method}
\label{Sol_MCC}
 \emph{MCC} method allots to those cells of $\boldsymbol{\chi}$ considering the lowest cost from $\boldsymbol{C}$. Firstly, the method allows the maximum permissible to the cell with the lowest per RB cost. Secondly, the amount of quantity and need is synthesized while crossing out the satisfied row(s) or column(s). 
 Either row or column is ruled out if both of them are satisfied concurrently. Thirdly, we inquire into the uncrossed-out cells which have the least unit cost and continue it till there is specifically one row or column is left uncrossed. The primary steps of the \emph{MCC} method are compiled as follows: 
 
\begin{itemize}
	\item[] \textbf{Step 1:} Distribute maximum permissible to the worthwhile cell of $\boldsymbol{\chi}$ which have the minimum cost found from  $\boldsymbol{C}$, and update the supply ($\boldsymbol{s}$) and demand ($\boldsymbol{d}$).
	\item[] \textbf{Step 2:} Continue \textbf{Step 1} till there is any demand that needs to be satisfied.
\end{itemize}

\subsubsection{\emph{MODI} Method for Finding an Optimal Solution}
 \label{Sol_MODI}
 The initial solution found from section \ref{Sol_MCC} is used as input in the \emph{MODI} method for finding an optimal solution. We need to augment an extra left-hand column and the top row (indicated by $x_u$ and $y_e$ respectively) with $\boldsymbol{C}$ whose values require to be calculated. The values are measured for all cells which have the corresponding allocation in $\boldsymbol{\chi}$ and shown as follows:
   \begin{equation}\label{Equ_MODI1}
   \begin{split}
    x_u+y_e=c_{u,e}, \forall \chi_{u,e}\ne \emptyset.
   \end{split}
   \end{equation}
  
Now we solve (\ref{Equ_MODI1}) to obtain all $x_u$ and $y_e$. If necessary then assign zero to one of the unknowns toward finding the solution. Next, evaluate for all the empty cells of $\boldsymbol{\chi}$ as follows:
 \begin{equation}\label{Equ_MODI2}
 \begin{split}
 k_{u,e}=c_{u,e}-x_u-y_e, \forall \chi_{u,e}= \emptyset.
 \end{split}
 \end{equation}

Now select $k_{u,e}$ corresponding to the most negative value and determine the stepping-stone path for that cell to know the reallocation amount to the cell. Next, allocate the maximum permissible to the empty cell of $\boldsymbol{\chi}$ corresponding to the selected $k_{u,e}$. $x_u$ and $y_e$ values for $\boldsymbol{C}$ and $\boldsymbol{\chi}$ must be recomputed with the help of (\ref{Equ_MODI1}) and a cost change for the empty cells of $\boldsymbol{\chi}$ need to be figured out using (\ref{Equ_MODI2}). A corresponding reallocation takes place just like the previous step and the process continues till there is a negative $k_{u,e}$. At the end of this repetitive process, we get the optimal allocation ($\boldsymbol{\chi}$). The \emph{MODI} method described above can be summed as follows:
  \begin{itemize}
  	\item[] \textbf{Step 1:} Develop a preliminary solution ($\boldsymbol{\chi}$) applying the \emph{MCC} method.
  	\item[] \textbf{Step 2:} For every row and column of $\boldsymbol{C}$, measure $x_u$ and $y_e$ by applying (\ref{Equ_MODI1}) to each cell of $\boldsymbol{\chi}$ that has an allocation. 
  	\item[] \textbf{Step 3:} For every corresponding empty cell of $\boldsymbol{\chi}$, calculate $k_{u,e}$ by applying (\ref{Equ_MODI2}).  
  	\item[] \textbf{Step 4:} Determine the stepping-stone path \cite{Charnes1954} from $\boldsymbol{\chi}$ corresponding to minimum $k_{u,e}$ that found in \textbf{Step 3}. 
  	\item[] \textbf{Step 5:} Based on the stepping-stone path found in \textbf{Step 4}, allocate the highest possible to the free cell of $\boldsymbol{\chi}$.
  	\item[] \textbf{Step 6:} Reiterate \textbf{Step 2} to \textbf{5} until all $k_{u,e}\ge 0$.
  \end{itemize}
  

\subsection{Low-Complexity Heuristic Algorithm for Solving Sub-Problem (\ref{Opt1_1})}
\label{Heu_Algo}
Though Algorithm \ref{Algo1} can solve the sub-problem (\ref{Opt1_1}) optimally, but computation time requires to solve it grows much faster as the size of the problem increase. Besides, the number of \emph{eMBB} UEs is large in reality, and we have a short period to resolve this kind of problem. Therefore, we need a faster and efficient heuristic algorithm, which may sacrifice optimality, to solve (\ref{Opt1_1}). Thus, we propose Algorithm \ref{Algo2} for solving  (\ref{Opt1_1}). At $t=1$, Algorithm \ref{Algo2} allocate resources equally to the \emph{eMBB} UEs. But, it allocates resources to \emph{eMBB} UEs in the rest of the time slots depending on the proportional loss of the previous time slot. In this way, Algorithm \ref{Algo2} can accommodate the EAR of \emph{eMBB} UEs in the long-run. The complexity of Algorithm \ref{Algo2} depends on $\mathcal{T}$ and $\mathcal{E}$.  

\begin{algorithm}
	\caption{Heuristic Algorithm for Solving (\ref{Opt1_1})}\label{Algo2}
	\begin{algorithmic}[1]
		\STATE $\mathbf{Initialization}$: $\varepsilon_1,\sigma_1,I_{max}$ and let $i=0$ 
		\STATE Solve problem (\ref{Opt1_1_2}) and obtain solution $\boldsymbol{\alpha}^{t,0}$
		\FOR {each $t\in \mathcal{T}$}
		\IF {t =1}
		\STATE Calculate $N_{RB}=\frac{|\mathcal{K}|}{|\mathcal{E}|}$
		\FOR {each $e\in \mathcal{E}$}
		\FOR {each $k=1\cdots N_{RB}$}
		\STATE $\alpha_{e,(e-1)*N_{RB}+k}^t=1$
		\ENDFOR
		\ENDFOR
		\ELSE
		\STATE Determine $r_{e,loss}^{t-1}$ and $r_{e,actual}^{t-1}$ for all $e\in \mathcal{E}$ by using ($\ref{Equ_loss}$) and ($\ref{Equ_actual}$) respectively
		\STATE Set $loc=0$
		\FOR {each $e\in \mathcal{E}$}
		\STATE Calculate $N_{RB}^e=\frac{r_{e,loss}^{t-1}}{\sum_{e'\in \mathcal{E}}r_{e',loss}^{t-1}}|\mathcal{K}|$
		\FOR {each $k=1\cdots N_{RB}^e$}
		\STATE $\alpha_{e,loc+k}^t=1$
		\ENDFOR
		\STATE Set $loc=loc+N_{RB}^t$ 
		\ENDFOR
		\ENDIF
		\ENDFOR
		\STATE Determine  $r_{e,actual}^{t}$ for all $e\in \mathcal{E}$ by using ($\ref{Equ_actual}$)
		\STATE Determine  $\mathbb{E}\bigg(\sum_{t=1}^{|\mathcal{T}|}r_{e,actual}^t\bigg)$ for all $e\in \mathcal{E}$ 
	\end{algorithmic} 
\end{algorithm}

\section{Numerical Analysis and Discussions}
 \label{Per_eva}
  In this section, we assess the proposed approach using
  comprehensive experimental analyses. Here, we compare our results with the results of the following state-of-the-art schedulers: 
  \begin{itemize}
  	\item $\textbf{PS}$\cite{Pedersen2017}: PS immediately overwrite part of the continuing \emph{eMBB} transmission with the sporadic \emph{uRLLC} traffic if there are not sufficient PRBs available. It chooses PRBs with the highest MCS that already been allotted to \emph{eMBB} UEs.   
  	\item $\textbf{MUPS}$\cite{Esswie2018}: In case of insufficient RBs, MUPS allocates PRBs to the \emph{uRLLC} UEs where they endure better channel quality depending on the CQI feedback.
  	\item $\textbf{RS}$: RS takes the RBs from the \emph{eMBB} UEs randomly in case of inadequate PRBs for supporting \emph{uRLLC} traffic.   
  	\item $\textbf{EDS}$: For supporting sporadic \emph{uRLLC} traffic, EDS offers the PRBs to this traffic after preempting PRBs equally from the \emph{eMBB} UEs in case of unavailable PRBs. 
  	\item $\textbf{MBS}$: gNB uses many to one matching game for snatching PRBs from \emph{eMBB} UEs for supporting \emph{uRLLC} traffic.
  \end{itemize}

  The main performance parameters are \emph{MEAR} and fairness \cite{Jain1984}  of the \emph{eMBB} UEs and defined as follows:
  \begin{equation}\label{Equ_MEAR}
  \begin{split}
   \mbox{MEAR} = \min \ \mathbb{E}\left(\sum_{t=1}^{|\mathcal{T}|}r_{e,actual}^t\right),\forall e\in \mathcal{E},
  \end{split}
  \end{equation}
  \begin{equation}\label{Equ_Fairness}
  \begin{split}
  \mbox{Fairness}=&\frac{\left(\sum_{e\in \mathcal{E}}\mathbb{E}\left(\sum_{t=1}^{|\mathcal{T}|}r_{e,actual}^t\right)\right)^2}{|\mathcal{E}|\cdot \sum_{e\in \mathcal{E}}\left(\sum_{t=1}^{|\mathcal{T}|}r_{e,actual}^t\right)^2}.
  \end{split}
  \end{equation}
In our scenario, we consider an area with a radius of $200$ m and gNB resides in the middle of the considered area. \emph{eMBB} and \emph{uRLLC} UEs are disseminated randomly in the coverage space. gNB works on a $10$ MHz licensed band for supporting the UEs in downlink mode. Every \emph{uRLLC} UE needs a single PRB for its service. Furthermore, gNB estimates path-loss for both \emph{eMBB} and \emph{uRLLC} UEs using a free space propagation model amidst Rayleigh fading. Table \ref{Table2} exhibits the significant parameters for this experiment. We use similar \emph{PSUM} parameters as of \cite{Zhang2017}. We realize the results of every approaches after taking $1,000$ runs. 

\begin{table}
	\centering
	\renewcommand{\arraystretch}{1.5}
	\caption{Summary of the simulation setup}\label{Table2}
	\begin{tabular}{|c|c|c|c|} \hline
		$\textbf{Symbol}$ &$\textbf{Value}$&$\textbf{Symbol}$&$\textbf{Value}$\\ \hline
		$|\mathcal{E}|$ & $10$& $|\mathcal{K}|$ & $50$\\ \hline
		$B$ & $180$ kHz  &$\epsilon$ & $0.01$\\ \hline
		$|\mathcal{T}|$ & $1000$  &$M$ & $8$\\ \hline
		$\Delta$ & $1$ms  &$\delta$ &$0.125$ ms\\ \hline
		$P_e,\forall e$ & $21$ dBm & $P_u,\forall u$ & $21$ dBm \\ \hline
		$I_{max}$ & $20$ & $N_0$ & $-114$ dBm \\ \hline
		$\sigma$ & \multicolumn{3}{c|}{$1,2,\cdots,10$}\\ \hline
		$L$ & \multicolumn{3}{c|}{$32,50,100,150,200$ bytes}\\ \hline
		$\mbox{eMBB traffic model}$ & \multicolumn{3}{c|}{Full buffer}\\ \hline
		$\sigma_{1}$ & $2$ & $\epsilon_{1}$ & $0.001$  \\ \hline
		$\eta$ & $0.7$ & $\zeta$ & $1.1$  \\ \hline
	\end{tabular}
	\renewcommand{\arraystretch}{1}
\end{table}

\begin{figure}
	\centering
	\includegraphics[width=0.5\textwidth]{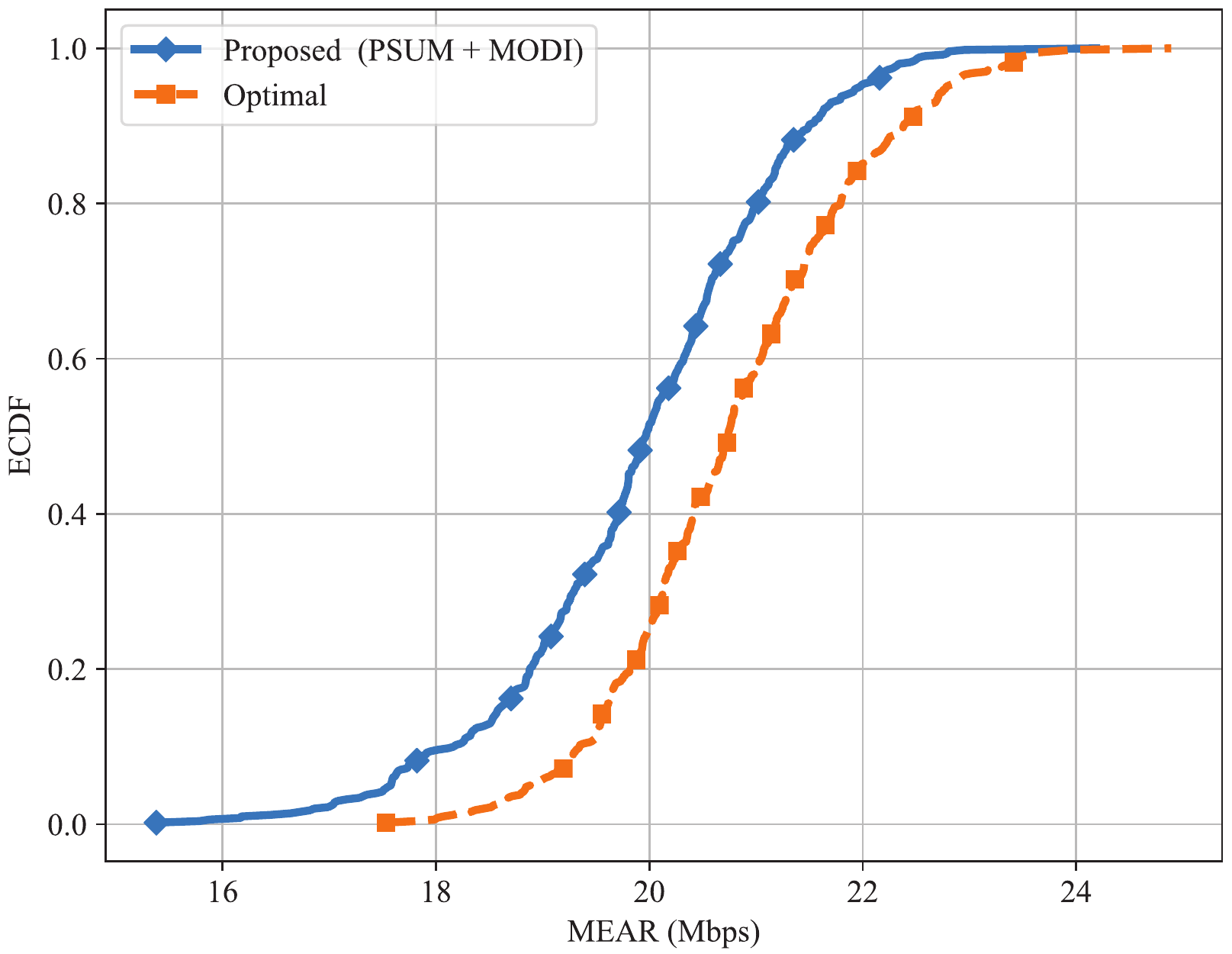}
	\caption{Comparison of \emph{MEAR} during $\mathcal{E}=4$ and single \emph{uRLLC} UE in every mini-slot when $L=32$ bytes.}\label{Fig_MAR_Opt}
\end{figure}

\begin{figure}
	\centering
	\includegraphics[width=0.5\textwidth]{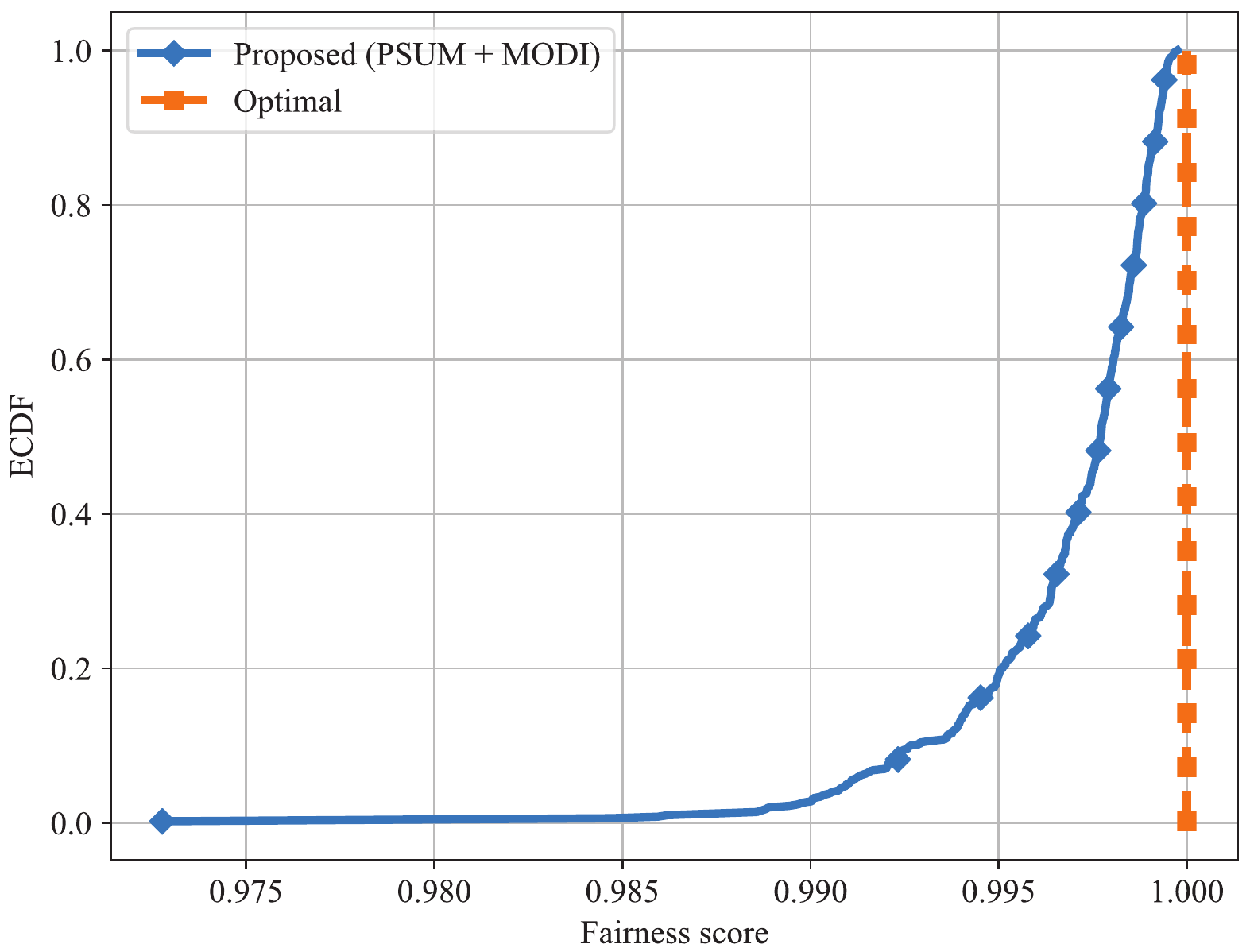}
	\caption{Comparison of fairness score when $\mathcal{E}=4$ and single \emph{uRLLC} UE in each mini-slot along with $L=32$ bytes.}\label{Fig_FN_Opt}
\end{figure}

A comparison of \emph{MEAR} and fairness scores are presented in Fig. \ref{Fig_MAR_Opt} and Fig. \ref{Fig_FN_Opt}, respectively, between the proposed (\emph{PSUM}+\emph{TM}) and the optimal value for a small network. Fig. \ref{Fig_MAR_Opt} shows the ECDF of \emph{MEAR} and the probability of \emph{MEAR} being at least $20$ Mbps are around $0.50$ and $0.70$, respectively, for the proposed and optimal methods, consequently. The optimality gap of average \emph{MEAR} for the proposed method is $4.20\%$ as represented in Fig. \ref{Fig_MAR_Opt}. Fig. \ref{Fig_FN_Opt} shows the ECDF of the fairness scores where the probability of the scores being $0.995$ at least is $0.80$ in the proposed method in comparison of being $1$ in the optimal mechanism. The optimality gap of the proposed method for the average fairness score is $0.32\%$ as exposed from Fig. \ref{Fig_FN_Opt}. 
       
\begin{figure}
	\centering
	\includegraphics[width=0.5\textwidth]{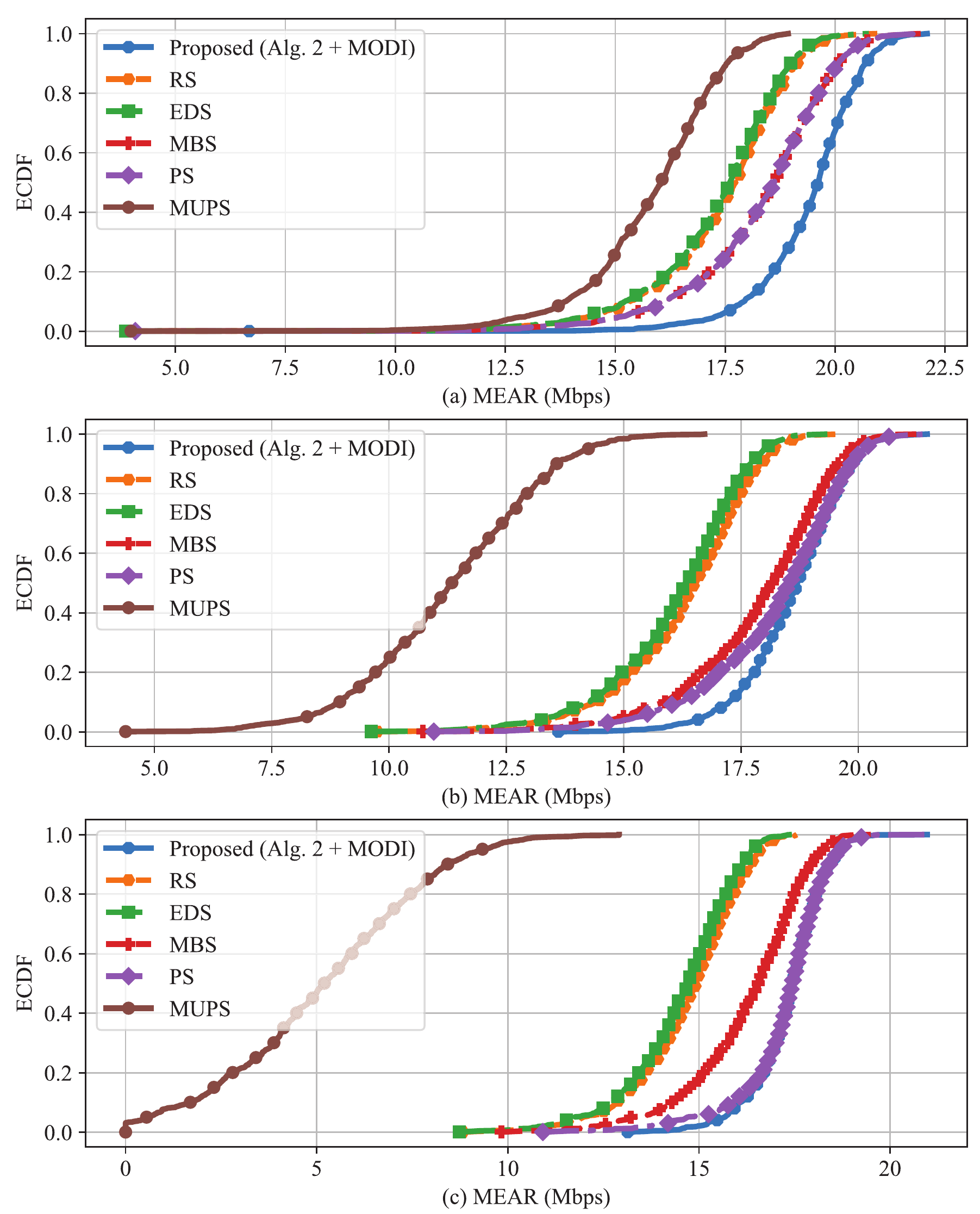}
	\caption{Comparison of \emph{MEAR} for (a) $\sigma=1$, (b) $\sigma=5$, and (c) $\sigma=10$, along with $L=32$ Bytes.}\label{Fig_MAR1}
\end{figure}

For growing \emph{uRLLC} arrivals, the ECDF of the \emph{MEAR} values is exhibited in Fig. \ref{Fig_MAR1}. Fig. \ref{Fig_MAR1} reveals the results that are preferred to those of the other considered methods. 
The probability of \emph{MEAR} values for being at least $18.0$ Mbps are 
$0.889$, $0.405$, $0.367$, $0.653$, $0.653$, and $0.052$ for the proposed, RS, EDS, MBS, PS, and MUPS methods, respectively, that are shown in Fig.  \ref{Fig_MAR1}(a). Fig.  \ref{Fig_MAR1}(b) reveals that the likelihood of \emph{MEAR} values for obtaining a minimum of $18.0$ Mbps are $0.736$, $0.089$, $0.050$, $0.541$, and $0.647$ for the proposed, RS, EDS, MBS, and PS methods, respectively, while the MUPS method can accommodate under $18$ Mbps in every case. Fig. \ref{Fig_MAR1}(c) shows that the proposed, MBS and PS methods provide a minimum \emph{MEAR} value of $18.0$ Mbps with a probability $0.231$, $0.089$, and $0.231$, respectively, while RS, EDS, and MUPS can produce less than $18$ Mbps for sure. Moreover, the \emph{MEAR} value decreases with the growing rate of $\sigma$ for all the methods because of the requirement of more RBs for the \emph{uRLLC} UEs as shown in Fig. \ref{Fig_MAR1}.  But, the increasing arrivals of \emph{uRLLC} traffic affect the MUPS method more as they require extra RBs from the distant \emph{eMBB} UEs. However, the performance gap between the proposed and PS method reduces with the increased arrival of \emph{uRLLC} traffic, as the PS scheme gets more chance to adjust the users with the higher expected achieved rate.      

\begin{figure}
	\centering
	\includegraphics[width=0.5\textwidth]{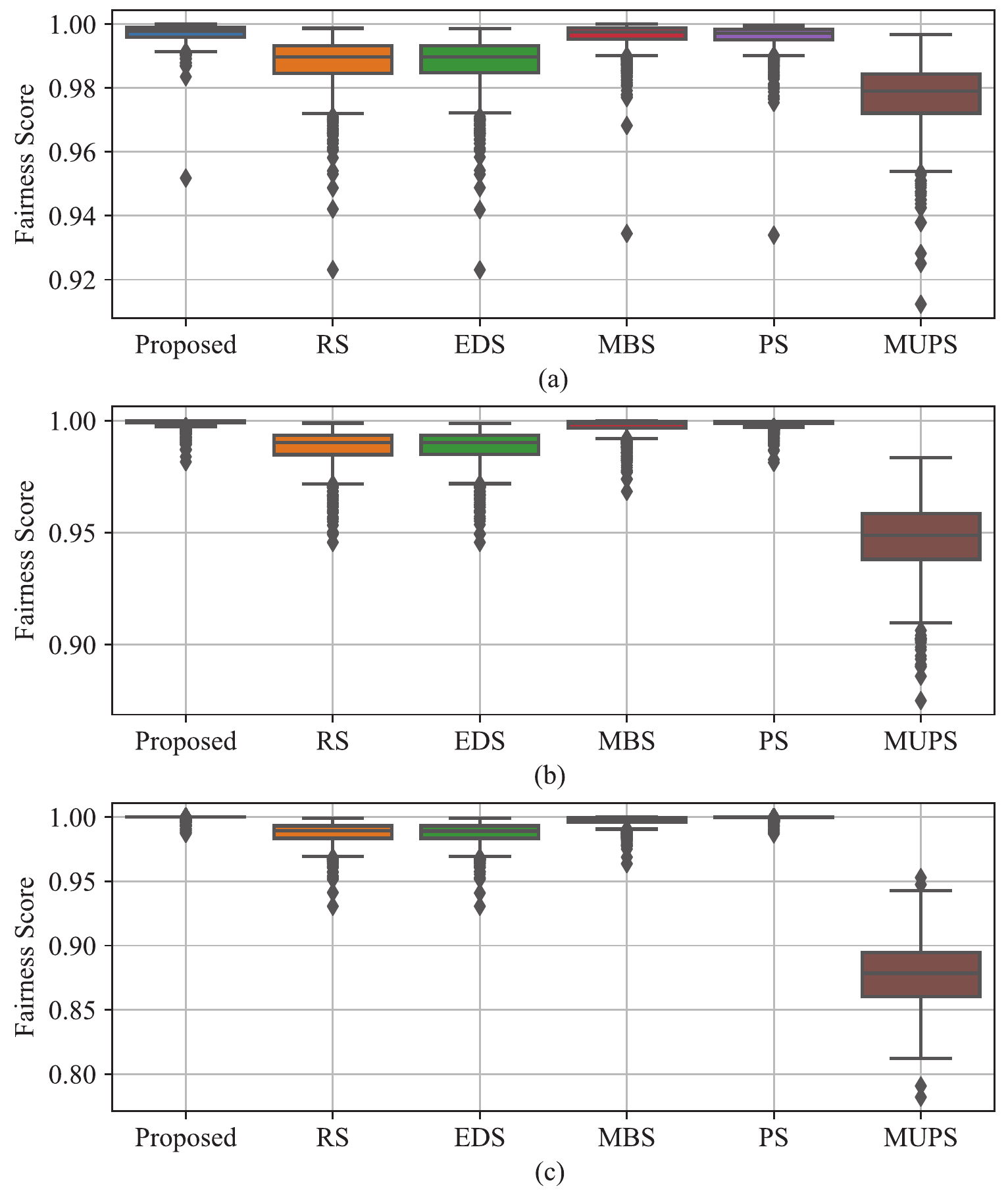}
	\caption{Comparison of fairness scores (a) $\sigma=1$, (b) $\sigma=5$, and (c) $\sigma=10$, along with $L=32$ Bytes.}\label{Fig_FN1}
\end{figure}

We compare the fairness scores among various methods with different values of $\sigma$ which is shown in Fig. \ref{Fig_FN1}. The scores originating from the proposed method are greater than or similar to that of others as indicated in Fig. \ref{Fig_FN1}. Fig. \ref{Fig_FN1}(a) reveals that the median of the scores for the proposed, RS, EDS, MBS, PS, and MUPS methods are $0.9977$, $0.9897$, $0.9897$, $0.9975$, $0.9972$, and $0.9789$, respectively. The similar scores are $0.9998$, $0.9902$, $0.9902$, $0.9987$, $0.9995$, $0.9488$, and $1.00$, $0.9891$, $0.9891$, $0.9985$, $0.9998$, $0.8784$ for the corresponding methods and are presented in Fig. \ref{Fig_FN1}(b) and \ref{Fig_FN1}(c), respectively. Moreover, the fairness scores increase for the Proposed, MBS and PS methods with the increasing value of $\sigma$ as it gets more chance to maximize the minimum achieved rate, whereas the same scores decrease with the increasing value of $\sigma$ for RS, EDS and MUPS as \emph{eMBB} UEs have more opportunity to be affected by the \emph{uRLLC} UEs.  

\begin{figure}
	\centering
	\includegraphics[width=0.5\textwidth]{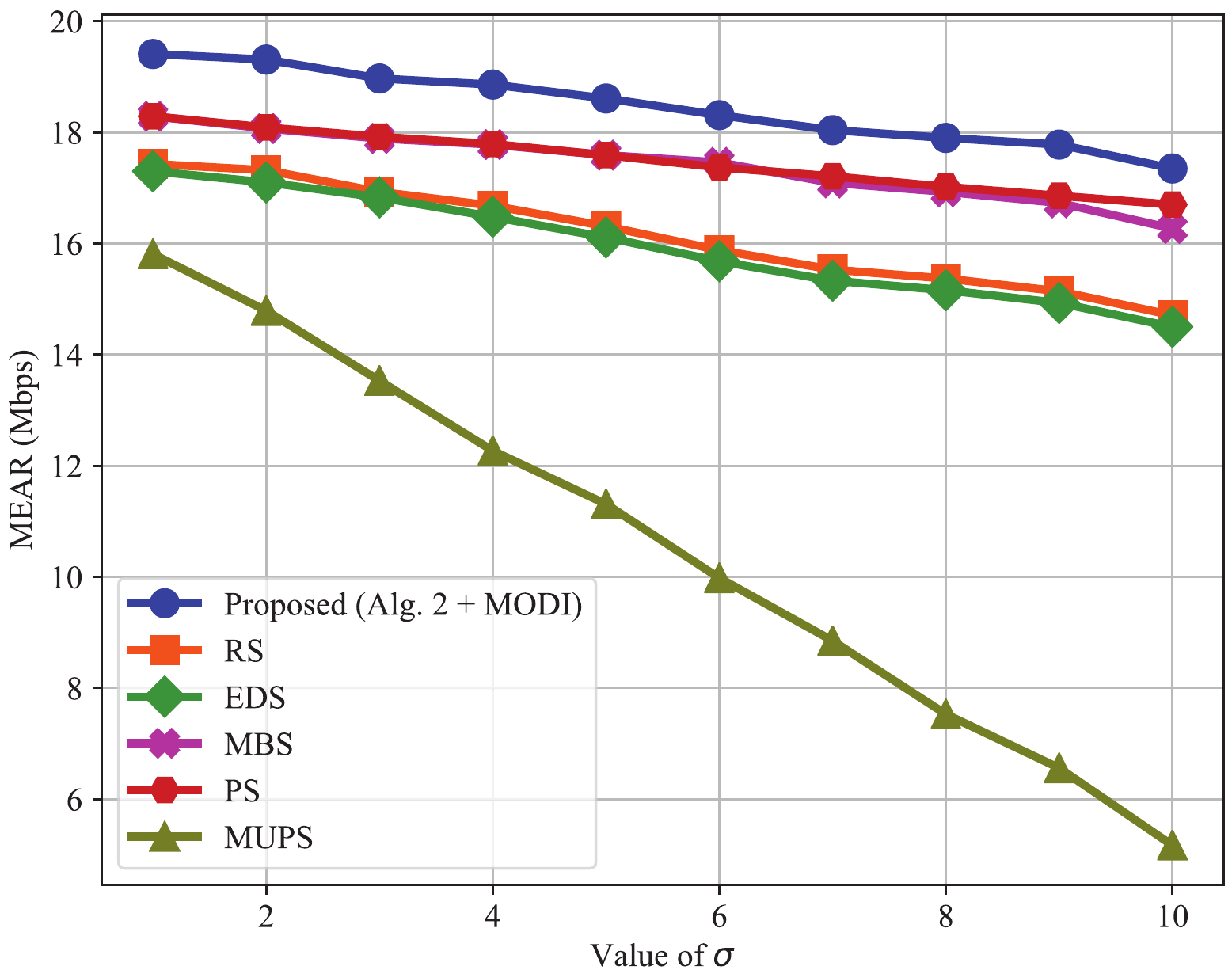}
	\caption{Comparison of average \emph{MEAR} with varying value of $\sigma$ and $L=32$ Bytes.}\label{Fig_MAR2}
\end{figure}

\begin{figure}
	\centering
	\includegraphics[width=0.5\textwidth]{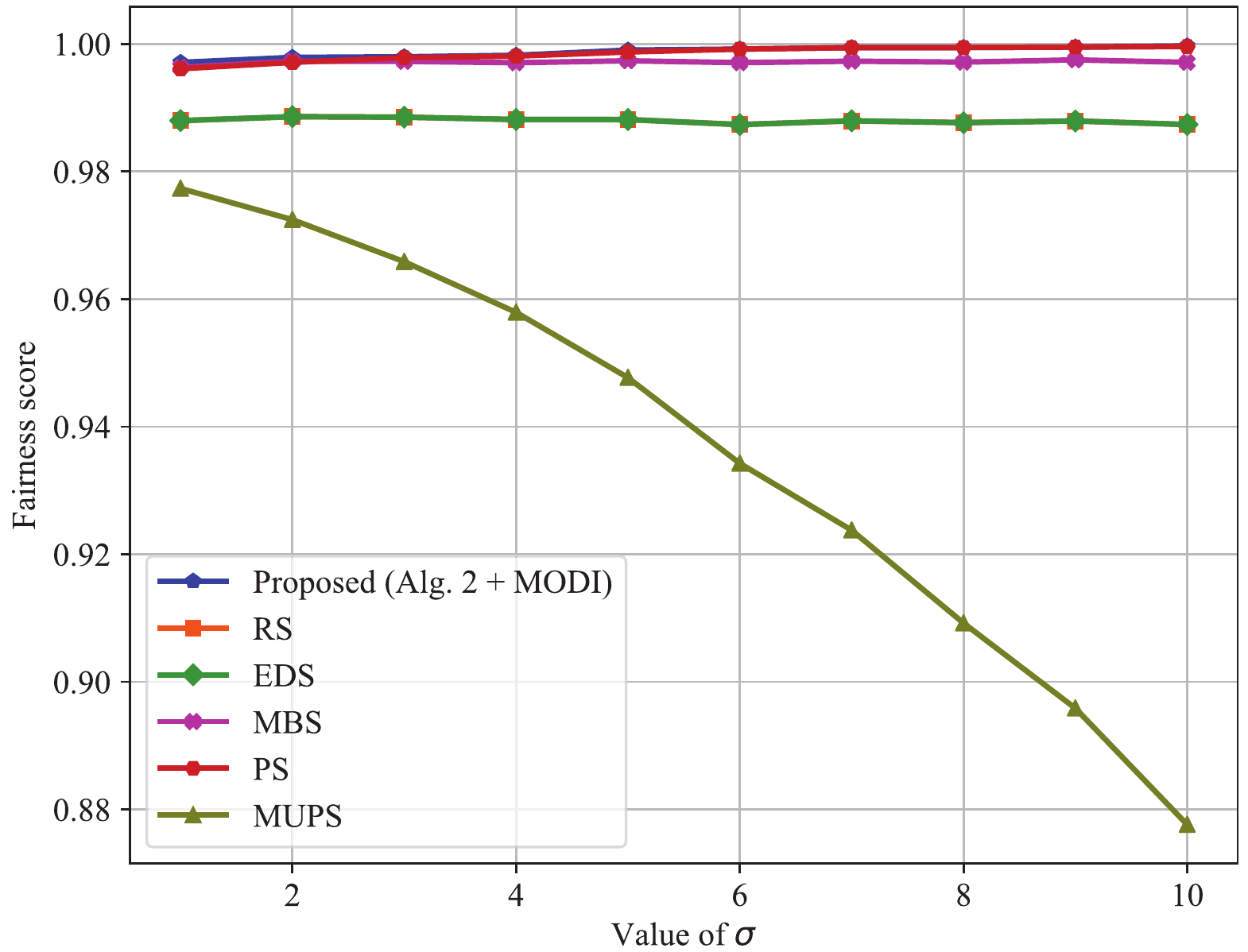}
	\caption{Comparison of fairness score with varying value of $\sigma$ and $L=32$ Bytes.}\label{Fig_FN2}
\end{figure}

Fig. \ref{Fig_MAR2} and \ref{Fig_FN2}, respectively, show the average \emph{MEAR} and fairness score for varying value of $\sigma$. In Fig. \ref{Fig_MAR2}, we find that our method overpasses other schemes for different rates of $\sigma$ in the case of average \emph{MEAR}. The figure also explicates that the average \emph{MEAR} is declining with the growing value of $\sigma$ due to the additional requirement of PRBs for extra \emph{uRLLC} traffic. Particularly, our method results $10.20\%$, $10.87\%$, $5.77\%$, $5.77\%$, and $18.55\%$ higher on average \emph{MEAR} than those of RS, EDS, MBS, PS, and MUPS, respectively, for $\sigma=1$. Moreover, similar values are $15.22\%$, $16.43\%$, $6.22\%$, $3.75\%$, and $70.20\%$ for $\sigma=10$. The average fairness score emerging from our method is bigger than or similar to other comparing methods for different values of $\sigma$ and shown in Fig. \ref{Fig_FN2}. Fig. \ref{Fig_FN2} also reveals that the $\sigma$ value has a negligible impact on the average score of the fairness in the Proposed, RS, EDS, MBS, PS methods, but it impacts inversely to the MUPS method more and more \emph{uRLLC} traffic choose same \emph{eMBB} UE for the PRBs. Moreover, the average fairness scores of the proposed method are similar to both MBS and PS methods. However, the proposed method treats \emph{eMBB} UEs $0.92\%$, $0.92\%$, and $1.92\%$ fairly than RS, EDS, and MUPS methods , respectively, when $\sigma=1$, whereas, the similar scores are $1.23\%$, $1.23\%$, and $12.21\%$, respectively, during $\sigma=10$.  

\begin{figure}
	\centering
	\includegraphics[width=0.5\textwidth]{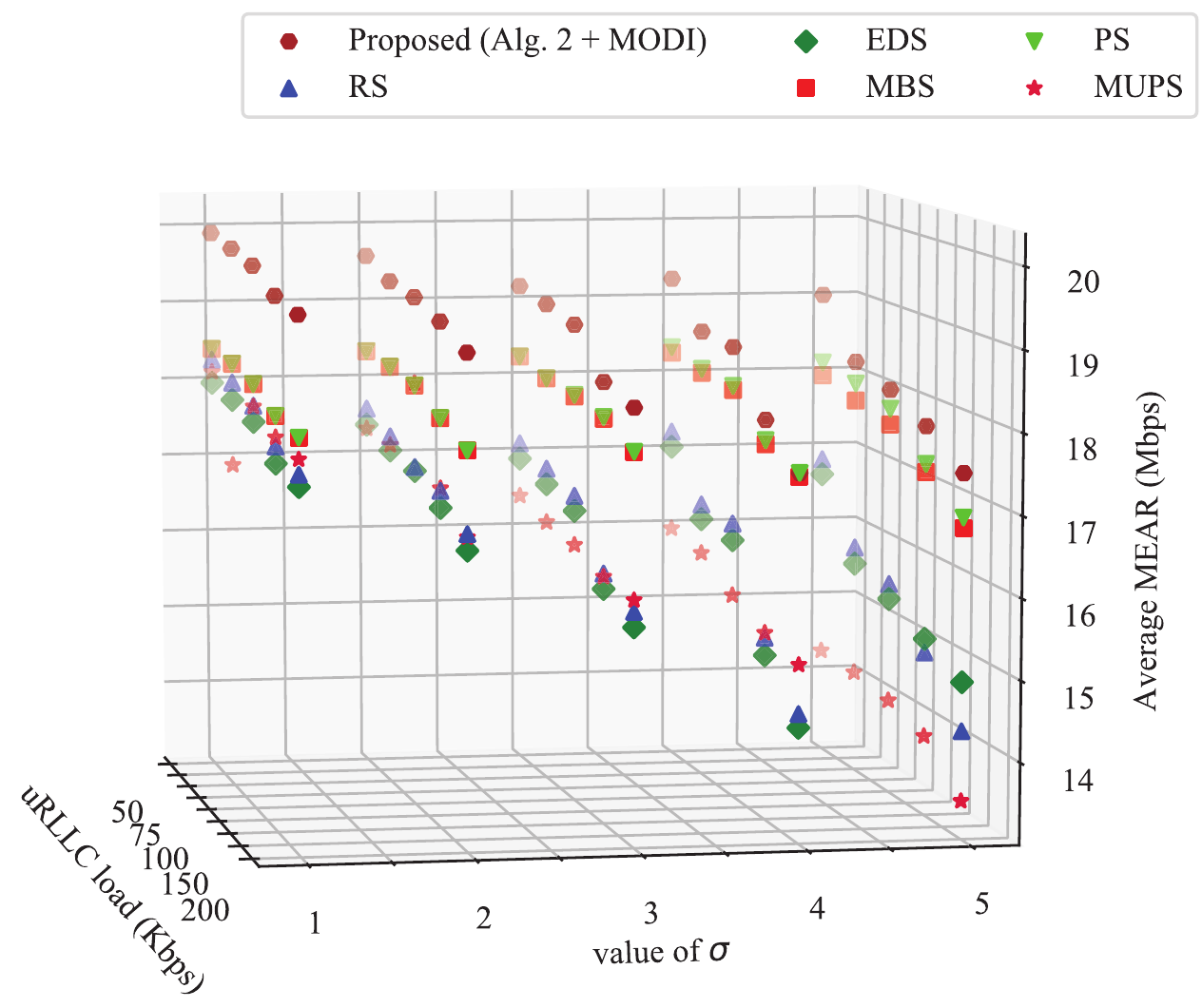}
	\caption{Comparison of average \emph{MEAR} with varying \emph{uRLLC} load and $\sigma$.}\label{Fig_MAR3}
\end{figure}

\begin{figure}
	\centering
	\includegraphics[width=0.5\textwidth]{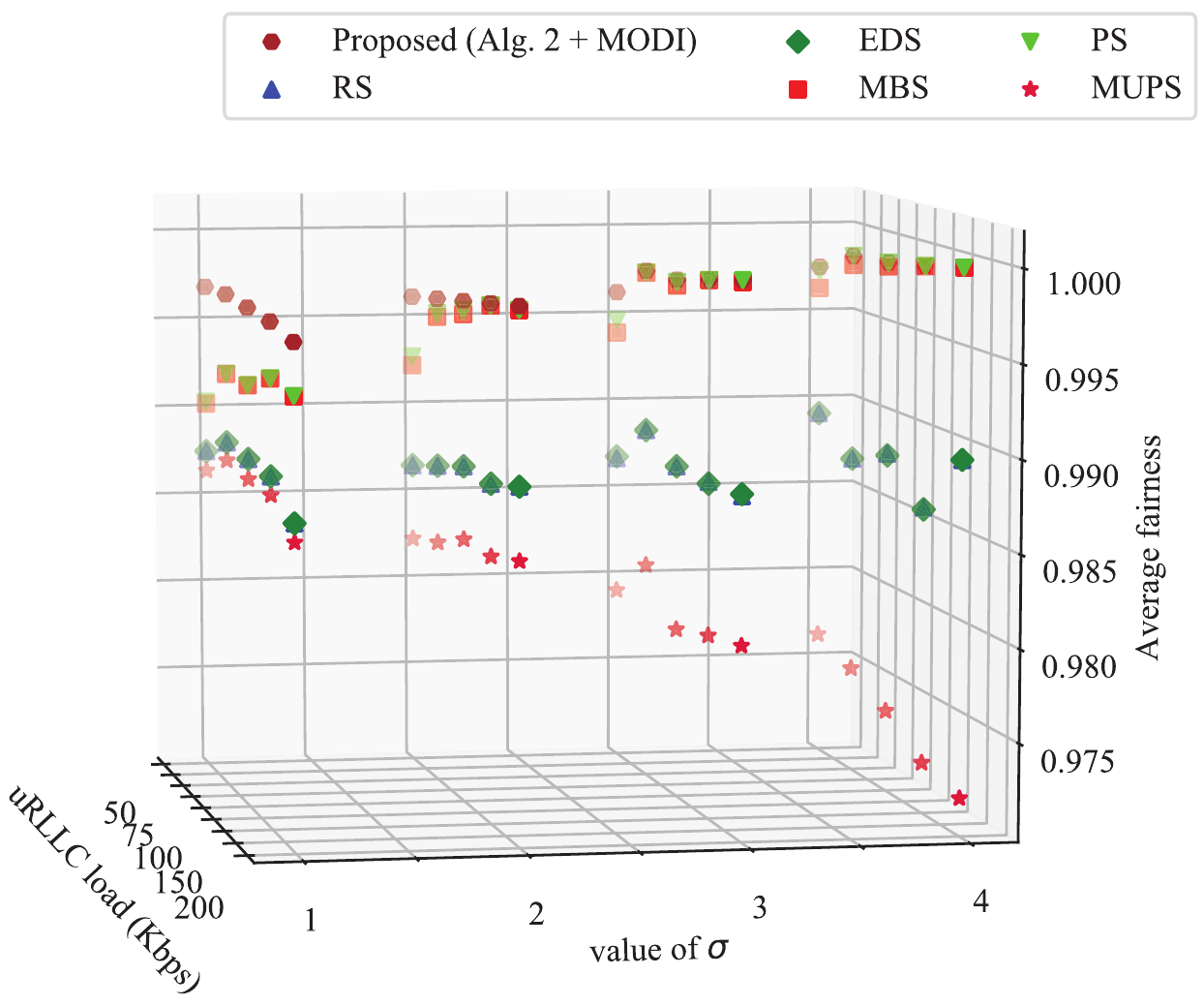}
	\caption{Comparison of average fairness score with varying \emph{uRLLC} load and $\sigma$.}\label{Fig_FN3}
\end{figure}

In Fig. \ref{Fig_MAR3}, we compare the average \emph{MEAR} of \emph{eMBB} UEs for considering varying \emph{uRLLC} load ($L$) and \emph{uRLLC} traffic ($\sigma$).  The \emph{MEAR} value of our method surpasses other concerned methods in every circumstance as revealed from Fig. \ref{Fig_MAR3}. The same figure also explicates that these values degrade when $L$ increases for varying  $\sigma$ as the system needs to allocate more PRBs to the \emph{uRLLC} UEs. 
Moreover, these values decrease with the increasing value of $\sigma$ for a fixed $L$, and also the same for increasing the value of $L$ with a fixed $\sigma$. In Fig. \ref{Fig_FN3}, we compare the average fairness score of \emph{eMBB} UEs for the different methods for changing the \emph{uRLLC} load ($L$) and \emph{uRLLC} traffic ($\sigma$). Fig. \ref{Fig_FN3} exposes that the fairness scores of our method are better than or at least similar to that of its' rivals. The figure also reveals that these scores decrease with an increasing $L$ for the lower value of $\sigma$. However, these scores increase with the increasing $L$ when $\sigma$ value is high. Moreover, for the MUPS method, these values decrease with the increasing value of $\sigma$ and $L$.    

\section{Conclusions}
\label{Conc}
In this paper, we have introduced a novel approach for coexisting \emph{uRLLC} and \emph{eMBB} traffic in the same radio resource for enabling 5G wireless systems. We have expressed the coexisting dilemma as a maximizing problem of the \emph{MEAR} value of \emph{eMBB} UEs meanwhile attending the \emph{uRLLC} traffic. We handle the problem with the help of the decomposition strategy. In every time slot, we resolve the resource scheduling sub-problem of \emph{eMBB} UEs using a \emph{PSUM} based algorithm, whereas the similar sub-problem of \emph{uRLLC} UEs is unraveled through optimal transportation model, namely \emph{MCC} and \emph{MODI} methods. For the efficient scheduling of PRBs among \emph{eMBB} UEs, we also present a heuristic algorithm. Our extensive simulation outcomes demonstrate a notable performance gain of the proposed approach over the baseline approaches in the considered indicators.





%

\end{document}